\documentclass{article}

\usepackage{arxiv}

\usepackage[utf8]{inputenc} % allow utf-8 input
\usepackage[T1]{fontenc}    % use 8-bit T1 fonts
\usepackage{hyperref}       % hyperlinks
\usepackage{url}            % simple URL typesetting
\usepackage{booktabs}       % professional-quality tables
\usepackage{amsfonts}       % blackboard math symbols
\usepackage{nicefrac}       % compact symbols for 1/2, etc.
\usepackage{microtype}      % microtypography
\usepackage{lipsum}		% Can be removed after putting your text content
\usepackage{graphicx}
\usepackage{natbib}
\usepackage{doi}
\usepackage{multirow} 
\usepackage{multicol}
\usepackage{subfigure}
\usepackage[flushleft]{threeparttable}
\usepackage{array}
\usepackage{mathtools}
\usepackage{tikz}

    \newcommand\Myperm[2][^n]{\prescript{#1\mkern-2.5mu}{}P_{#2}}
    \newcommand\Mycomb[2][^n]{\prescript{#1\mkern-0.5mu}{}C_{#2}}

\title{SubLock: Sub-Circuit Replacement based Input Dependent Key-based Logic Locking for Robust IP Protection}

\author{ {Vijaypal Singh Rathor}\\
	Department of CSE\\
	PDPM IIITDM Jabalpur\\
	India, 482005 \\
	\texttt{vrathor@iiitdmj.ac.in} \\
	\And
	{Munesh Singh} \\
	Department of CSE\\
	PDPM IIITDM Jabalpur\\
	India, 482005 \\
	\texttt{munesh.singh@iiitdmj.ac.in} \\
	  \AND
	{Kshira Sagar Sahoo} \\
	Department of Computing Science\\
	Ume\r{a} University, Ume\r{a}\\
	Sweden, 901 87 \\
	\texttt{kshirasagar12@gmail.com} \\
        \And
	{Saraju P. Mohanty} \\
	Department of CSE\\
	University of North Texas, Denton\\
	TX, USA, 76207 \\
	\texttt{saraju.mohanty@unt.edu} \\
}

\renewcommand{\shorttitle}{\textit{arXiv} Template}

\hypersetup{
pdftitle={A template for the arxiv style},
pdfsubject={q-bio.NC, q-bio.QM},
pdfauthor={David S.~Hippocampus, Elias D.~Striatum},
pdfkeywords={First keyword, Second keyword, More},
}

\begin{document}
\maketitle

\begin{abstract}
  Intellectual Property (IP) piracy, overbuilding, reverse engineering, and hardware Trojan are serious security concerns during integrated circuit (IC) development. Logic locking has proven to be a solid defence for mitigating these threats. The existing logic locking techniques are vulnerable to SAT-based attacks. However, several SAT-resistant logic locking methods are reported; they require significant overhead. This paper proposes a novel input dependent key-based logic locking (IDKLL) that effectively prevents SAT-based attacks with low overhead. We first introduce a novel idea of IDKLL, where a design is locked such that it functions correctly for all input patterns only when their corresponding valid key sequences are applied. In contrast to conventional logic locking, the proposed IDKLL method uses multiple key sequences (instead of a single key sequence) as a valid key that provides correct functionality for all inputs. Further, we propose a sub-circuit replacement based IDKLL approach called SubLock that locks the design by replacing the original sub-circuitry with the corresponding IDKLL based locked circuit to prevent SAT attack with low overhead. The experimental evaluation on ISCAS benchmarks shows that the proposed SubLock mitigates the SAT attack with high security and reduced overhead over the well-known existing methods.
\end{abstract}

% keywords can be removed

\section{Introduction}
  Recent advancement in technology ($i.e.$, artificial intelligence, Internet of Thing (IoT) and cyber-physical systems) encourages using high-functioning, automated and intelligent electronic devices, specifically IC, for various mission-critical applications such as financial, national defence, health care, transportation, and energy. Security is a critical concern in all these mission-critical applications. Due to the unaffordable cost of constructing and maintaining a foundry with advanced fabrication capabilities, semiconductor industries are becoming fabless. Further, critical time-to-market forces companies to use third-party intellectual property (IP) blocks to design an integrated circuit (IC). Due to these economic and timing constraints, outsourcing from third parties is unavoidable in IC development. The involvement of several untrusted third parties and people in IC development makes the IC vulnerable to various hardware-based attacks such as IP piracy, overbuilding, reverse engineering (RE), and hardware Trojan (HT) insertion \cite{rostami2014primer}, \cite{xiao2016hardware}. These attacks pose serious security concerns during the IC life cycle as they can compromise the whole system's security. For example, the insertion of hardware Trojan not only deviates IC from its normal function but also can leak sensitive information during infield operation \cite{bhunia2014hardware}. Further, as a result of these supply chain assaults, the IC industries lose billions of dollars every year \cite{Est2}, \cite{SEMI}.

   \begin{figure}[h!]    \centering
     \includegraphics[width=0.9\columnwidth]{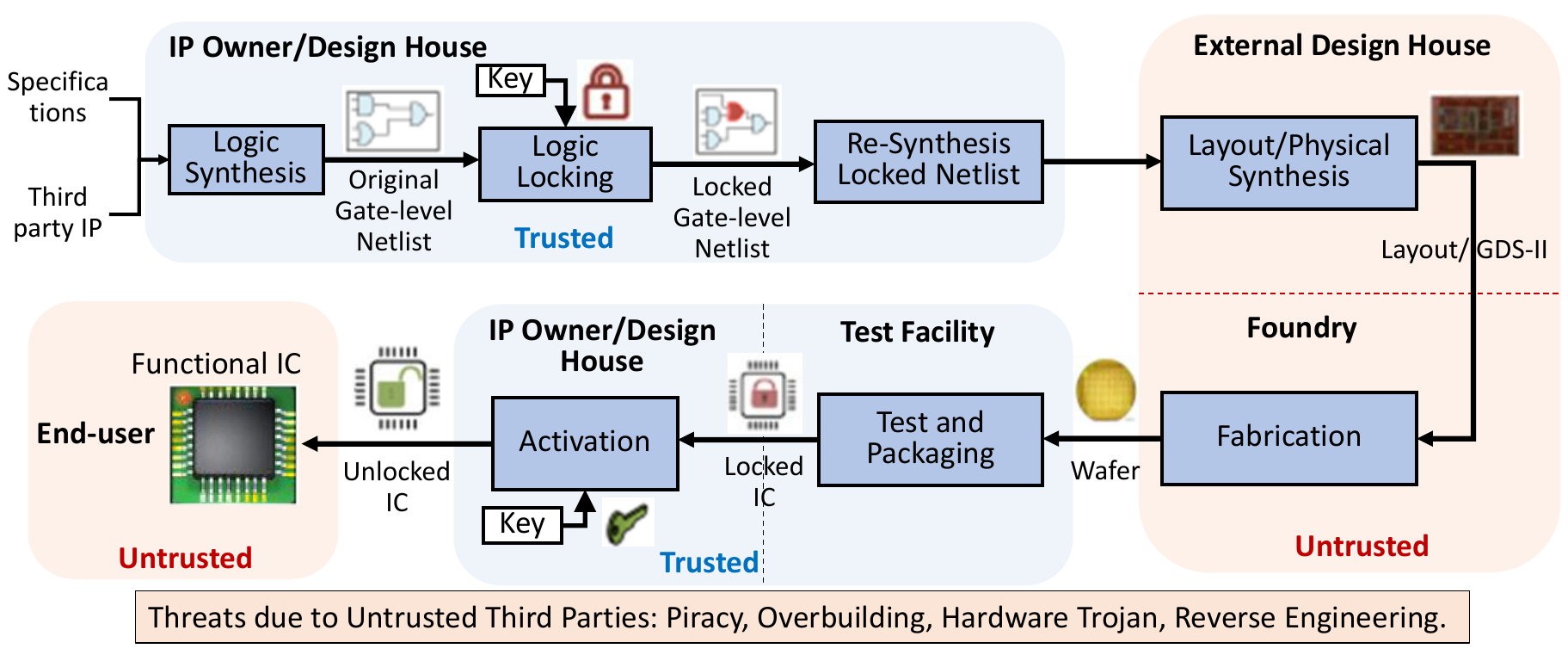}
      \caption{Logic Locking in IC development. Applying logic locking for mitigating various threats (i.e., IP Piracy, Overbuilding, Hardware Trojan and Reverse Engineering) during the IC life cycle.}
    \label{fig0}
  \end{figure}
 
  To mitigate these attacks, different design-based defence techniques such as removing rare-triggered nets \cite{salmani2012novel}, \cite{rathor2018novel}, HT detection \cite{narasimhan2013hardware}, \cite{salmani2017cotd}, logic locking/obfuscation \cite{roy2010ending} \cite{rajendran2015fault}, \cite{yasin2016improving}, logic camouflaging \cite{cocchi2014circuit}, \cite{Rathor2017}, etc. are reported in the literature to mitigates these threats. In the last five years, logic locking has emerged as the most effective and prominent method to thwart these attacks during the IC life cycle. Logic locking locks/obfuscates the design functionality by embedding a secret key sequence during the IC development, as shown in Figure \ref{fig0}. The design provides correct functionality only when a valid secret key sequence is applied. The correct key sequence is stored in on-chip tamper-evident memory, which is inaccessible to the attacker \cite{zhang2016practical}, \cite{rajendran2012logic}.

  The security of the logic locking techniques mainly depends on the secrecy of the key. Since the attacker always has access to the unlocked IC from the open market, he/she reveals the correct key by applying different attack mechanisms to the locked IC. In the last six years, boolean satisfiability (SAT) has emerged as the most effective attack that can compromise the security of a logic locking technique within a few minutes, even using a large key size. This attack iteratively eliminates the wrong keys by applying the distinguishing input patterns (DIPs) and identifying the correct key \cite{subramanyan2015evaluating},
  \cite{el2015integrated}
  . To mitigate the SAT attack, several SAT resilience logic locking techniques have been reported in the literature such as Anti-SAT block (ASB) \cite{xie2016mitigating}, \cite{xie2018anti}, SARLock \cite{yasin2016sarlock}, stripped functionality logic locking (SFLL) \cite{yasin2017provably}, cascaded locking (CAS-Lock) \cite{shakya2020cas} and Strong Anti-SAT (SAS) \cite{liu2020strong}. These methods either vulnerable to other attacks such as removal \cite{yasin2017security}, \cite{yasin2018removal}, AppSAT \cite{shamsi2017appsat}, Bypass \cite{xu2017novel}, Functional Analysis (FALL) \cite{sirone2020functional}, Identify flip signal (IFS) attack \cite{sengupta2021breaking} or provide trade-off between security and effectiveness with large overhead. However, recently an input dependent key-based logic locking (IDKLL) method called GateLock to prevent the SAT attacks \cite{10375951}. This method may not be vulnerable to structural analysis attacks due to using fixed structures for IDKLL gates. This paper proposes a new lightweight IDKLL based SAT resilient method that can effectively mitigate SAT and structural analysis attacks and provides robust IP protection. The contributions of this paper are presented in the next section.

  \section {Contributions of This Paper}
       This section first presents the problem being addressed in this paper, the proposed solution, followed by the novelty and significance of the proposed solution.
     
     \subsection{Problem Addressed in this Paper}
      The IC is vulnerable to hardware-based attacks, $i.e.$, piracy, overbuilding, RE, HT, etc. Though several logic locking techniques attempt to mitigate these attacks, they are vulnerable to SAT-based attacks and require significant design overhead. Most of the existing SAT-resistant logic locking techniques are vulnerable to SAT variants such as App-SAT, Bypass, removal, and FALL and require high implementation costs. Further, the recent methods such as CAS-Lock and SAS also provide the trade-off between SAT and other attacks and require a large overhead. Further, to the best of our knowledge, none of the existing SAT-resistant logic locking techniques has focused on preventing SAT attacks instead of increasing SAT attack time or DIPs. This paper considers preventing SAT-based attacks completely with low overhead and providing effective protection during the IC life cycle.
      
     \subsection{Solution Proposed in this Paper}
     This paper proposes a novel lightweight sub-circuit replacement based input dependent key-based logic locking (IDKLL) technique to prevent SAT-based attacks. The proposed method uses multiple key sequences as the correct key to unlock the design. The IDKLL divides the input patterns into multiple sets and uses a different key sequence as a valid key for each set of inputs. This means a separate key sequence is used to unlock the circuit for a specific set of inputs. None of the key sequences exists out of total key space that can achieve correct functionality for all the input patterns. Therefore, the design functions correctly for all sets of inputs only when their respective correct key sequences are applied. Further, in the proposed methods, design functionality is locked by replacing the original sub-circuits with their corresponding IDKLL based locked circuits. Due to sub-circuit replacement, it is hard to apply structural analysis attacks.
            
   %   \begin{enumerate}
   %     \item Presents analysis and the limitations of existing logic locking techniques.
   %      
   %     \item We introduce a novel idea of input dependent key-based logic locking to prevent the SAT attack. We also propose using LUT-based tamper-proof memory to achieve input dependent key-based logic locking.
   %           
   %     \item Further, a lightweight sub-circuit replacement based IDKLL is proposed that locks the design by replacing the existing logic with the corresponding IDKLL based locked circuitry to prevent SAT-based attacks. 
   %          
   %     \item Present a quantitative security analysis of IDKLL against the brute force attack. Also, show that our method tremendously increases the security of logic locking against brute force attacks over the existing methods. 
   %  \end{enumerate}
     
 \subsection{Novelty and Significance of the Solution}
   Unlike existing techniques, the proposed IDKLL based technique utilizes multiple key sequences (instead of a single key sequence) as a valid key to lock/unlock the design. Design can provide correct functionality for all inputs only when their respective correct key sequence is applied. Instead of storing single key sequences in normal tamper-proof memory, the proposed IDKLL uses the LUT-based tamper-proof memory to store and retrieve the corresponding valid key sequences to the applied inputs. In the proposed method, the SAT attack eliminates all the key sequences as all the key sequences are individually incorrect. Thus SAT attack failed to identify the correct key sequences. The existing techniques suffer from a trade-off between SAT and effectiveness and other attacks and require a large overhead. In contrast, the proposed IDKLL provides high security against SAT attacks while providing effective protection against other attacks such as removal \cite{yasin2018removal}, App-SAT \cite{shamsi2017appsat}, and Bypass \cite{xu2017novel}. The quantitative security analysis of the proposed IDKLL shows that the attacker has to require more than brute force attempts to decipher the correct key sequences. Further, the experimental evaluation on ISCAS and ITC benchmarks shows that the proposed technique effectively prevents SAT-based attacks with low overhead over the recent SAT-resistant logic locking. 

  The rest of the paper is organized as follows: Section \ref{related} analyses the existing SAT-resistant logic locking techniques. Section \ref{Prop-1} presents a novel concept of input dependent key-based logic locking to prevent SAT attacks. Further, Section \ref{Prop_light} presents a lightweight sub-circuit replacement based IDKLL method followed by its quantitative security analysis. The experimental evaluation and comparative analysis of our technique are presented in Section \ref{exp}. The conclusion of the paper is provided in Section \ref{con}.

\section{Related Prior Works} \label{related}
%\section{SAT Resistant Logic Locking: An Analysis} \label{related}
  The researchers have proposed various logic locking techniques which either insert the additional key gates like XOR/XNOR \cite{roy2010ending}, \cite{rajendran2012logic}, AND/OR \cite{dupuis2014novel}, MUX \cite{rajendran2015fault} or replaces the existing logic with corresponding new locked logic \cite{juretus2016reducing}, LUTs \cite{liu2014embedded}. However, several attacks such as key-sensitization \cite{rajendran2012security}, \cite{yasin2016improving}, logic cone analysis \cite{lee2015improving}, hill climbing\cite{plaza2015solving}, SAT \cite{subramanyan2015evaluating}, \cite{el2015integrated} are reported to compromise the security above logic locking methods. Locking a design using an interference graph-based method and inserting an additional AES module in the locked design improves the security against these attacks \cite{yasin2016improving}. However, the interference graph-based method is topology dependent and vulnerable to logic cone analysis attacks. Also, AES insertion causes a large design overhead \cite{lee2015improving}. A lightweight and topology independent key-gate replacement based logic locking is reported in \cite{rathor2019lightweight} to neutralize the key-sensitization and logic cone analysis attacks.
     
   The SAT-resistant logic locking techniques such as SARLock \cite{yasin2016sarlock}, Anti-SAT block \cite{xie2016mitigating}, and AND Tree insertion \cite{li2016provably} are reported to mitigate the SAT attack with reduced overhead. The AND tree insertion-based approach is vulnerable to sensitization-guided SAT attack \cite{yasin2018removal}. Whereas, the SARLock and Anti-SAT are vulnerable to removal \cite{yasin2017security}, \cite{yasin2018removal}, App-SAT \cite{shamsi2017appsat}, double DIP \cite{shen2017double}, and Bypass \cite{xu2017novel} attacks. Moreover, the security of SARALock and Anti-SAT block can also be compromised using SAT-based Signature \cite{shen2019sigattack} and Bit-flipping \cite{shen2018sat} attacks. Similar to the removal attacks, these attacks also separate the SAT-resistant block from the locked circuit. 
  %   The effective integration and obfuscation of the SAT-resistant blocks in the locked circuit can mitigate these attacks. 
   Although, the security of the Anti-SAT block against removal attack is increased \cite{xie2018anti} by obfuscating it using LUT-based design withholding and wire entanglement approach \cite{khaleghi2015ic}. The use of these LUT-based approaches for obfuscation introduces a large design overhead \cite{khaleghi2015ic}. A lightweight Anti-SAT design and obfuscation technique is reported in \cite{RATHOR2020Anti-SAT} to reduce the design overhead and increase the security against removal attacks. This technique increases security with reduced design overhead by constructing the Anti-SAT block using existing circuitry.
        
   Besides the above, a tenacious and traceless logic locking called TTlock is reported in \cite{yasin2017lock} where the SARLock is re-architected to mitigate the SAT and removal attacks simultaneously. The TTLock is an extension of SARLock where the original netlist is modified to protect a secret input pattern such that the output of the original and modified netlist differ only for one input pattern. Similar to SARLock and Ant-SAT, TTLock is also vulnerable to App-SAT \cite{shamsi2017appsat}, double DIP \cite{shen2017double} and Bypass \cite{xu2017novel} attacks. These attacks mainly compromise the security of the above SAT-resistant methods due to their low output corruptibility. Though it is reported that output corruptibility of Anti-SAT can be increased \cite{yasin2018removal}, \cite{xu2017novel}, increasing the output computability may decrease the effectiveness of the logic locking against SAT attack \cite{rajendran2015fault}, \cite{xu2017novel}, \cite{yasin2018removal}. 
     
   In order to mitigate the App-SAT, Double DIP and Bypass attacks, an extension of TTLock called stripped functionality logic locking (SFLL) has also been proposed in \cite{yasin2017provably}. The SFLL selects multiple protected input patterns to increase the output corruptibility and increase the security against App-SAT, Bypass and Double DIP attacks. However, increasing the number of protected patterns increases the corruptibility; it decreases SAT iterations and increases the required overhead \cite{liu2020strong}, \cite{shakya2020cas}. Thus, SFLL creates the trade-off between security and effectiveness \cite{liu2020strong}. Besides, the SFLL is also proven vulnerable to Functional Analysis (FALL) attack as reported in \cite{sirone2020functional}. An attack framework has been proposed in  \cite{yang2019stripped} that breaks the SFLL by exploiting structural traces left in the locked design.

     \begin{table*}[t!]     \centering
       \caption{SUMMARY OF DIFFERENT SAT RESISTANT LOGIC LOCKING TECHNIQUES ALONG WITH THE IDENTIFIED LIMITATIONS} 
        \label{Gap-Tab1}
        {\small
          \begin{tabular}{|p{3cm}|p{5.2cm}|p{5cm}|}   \hline    
          {\bf Techniques} & {\bf Objectives}  & {\bf Limitations} \\ \hline
                 
          Strong logic locking (SLL) \cite{yasin2016improving} & Inserts non-mutable key-gates along with AES circuit to thwart sensitization and SAT attacks & Large overhead due to insertion of AES circuit, vulnerable to cone-based attack  \\ \hline
   
          SARLock \cite{yasin2016sarlock}, and Anti-SAT \cite{xie2016mitigating}, \cite{xie2018anti} & Provide good security against SAT attack with reduced overhead over AES & Vulnerable to Removal, App-SAT, double DIP and Bypass attacks\\ \hline
      
          Tenacious and Traceless logic locking (TTLock) \cite{yasin2017lock} & Provides the security against SAT and Removal attacks simultaneously & Vulnerable to App-SAT, double DIP and Bypass attacks\\ \hline
      
          Stripped Functionality  logic locking (SFLL) \cite{yasin2017provably} & Provides the security against SAT, App-SAT, Bypass and Removal  attacks simultaneously & Increased overhead, vulnerable to FALL and structural attacks as mentioned in \cite{yang2019stripped}\\ \hline  
        
          Cascaded Locking (CAS-Lock) and Mirror CAS (M-CAS) \cite{shakya2020cas} & Provide security against SAT, App-SAT, Bypass, FALL, and Removal attacks simultaneously with low overhead & CAS/M-CAS vulnerable to IFS/MKBM-SAT and provide a trade-off between SAT and removal, and require large overhead \\ \hline  
        
          Strong Anti-SAT (SAS) \cite{liu2020strong}  & Provides the security against SAT, App-SAT, Bypass and Removal and other attacks simultaneously & Increased overhead, require additional overhead for obfuscation otherwise may be vulnerable to removal/IFS-SAT attacks \\ \hline  
          
         Proposed IDKLL based SubLock Method & Effectively mitigate SAT and its known variants such as App-SAT, Bypass and Removal simultaneously & Increased memory size due to storing multiple valid key sequences in LUT based temper-proof memory  \\ \hline   
         
       \end{tabular}}
      \end{table*}

   However, a CAS-Lock \cite{shakya2020cas}, and strong Anti-SAT called SAS \cite{liu2020strong} based logic locking methods are proposed to mitigate the threats of App-SAT, Bypass and simultaneously ensure the effectiveness against SAT attack. The CAS-Lock basically adopts the merits of SARlock, and Anti-SAT uses cascaded key controlled AND/OR gates. But it is found that CAS-Lock is vulnerable to removal attacks. Therefore, the authors \cite{shakya2020cas} have mirrored the CAS-Lock called M-CAS in the original design to mitigate the removal attacks at the increased design overhead. Although M-CAS increase the security against removal, the design becomes vulnerable to SAT attack. The authors in \cite{sengupta2021breaking} have proposed IFS/Key-bit mapping \& SAT called IFS-SAT/KBM-SAT that exploits the structural traces such as single point function and breaks the CAS and M-CAS locking. Another side, SAS also adopts the merits of Anti-SAT and SFLL, where it introduces the error in the design for the selected set of input minterms called critical minterms. Though SAS based locking achieves high effectiveness without compromising security against SAT attacks over SFLL, it increases design overhead significantly. Further, it will also require additional design overhead for obfuscating and integrating the SAS block with standard logic locking to thwart removal and other attacks. The summary of the different related SAT-resistant logic locking techniques and their identified limitations is presented in Table \ref{Gap-Tab1}.  
      
     \begin{table*}[h!]
       \centering
        \caption{Comparison of effectiveness of proposed and existing methods to mitigates different attacks. Here, ``$\checkmark$" and ``$\times$" denote a method mitigates and does not mitigate the specified attack respectively. The ``$-$" denotes the partial mitigation of the attack.}
         \label{Tabcomp}
         {\footnotesize
         \begin{tabular}{|c|c|c|c|c|c|c|c|c|}   \hline
          \multirow{2}{*}{Methods} &  \multicolumn{8}{c|}{Attacks} \\ \cline{2-9}         
           			& 	SAT  			&	APP-SAT		 & 	Bypass 		 &	Double DIP 	 	&	Removal 		&	FALL 		 	&	SFLL-Unlocked	&	IFS/KBM 		\\ \hline
           SARLock		& 	$\checkmark$	& 	$\times$ 	 &	$\times$ 	 &	$\times$ 	 	&	$\times$	 	& 	$\checkmark$ 	& 	$\checkmark$	& 	$\checkmark$  	\\ \hline
           Anti-SAT 	&	$\checkmark$ 	&	$\times$ 	 &	$\times$ 	 &	$\times$ 		&	$\times$ 		&	$\checkmark$	& 	$\checkmark$ 	&	$\times$ 		\\ \hline
           TTL 		&	$\checkmark$ 	&	$\times$ 	 &	$\times$	 & 	$\times$ 		&	$\checkmark$	& 	$\times$ 		&	$\times$ 		&	$\checkmark$    \\ \hline
           SFLL		& 	$\checkmark$ 	&	$\checkmark$ & 	$\checkmark$ &	$\checkmark$	& 	$\checkmark$	& 	$\times$ 		&	$\times$		& 	$\checkmark$	\\ \hline
           CAS			& 	$\checkmark$ 	&	$\checkmark$ & 	$\checkmark$ &	$\checkmark$	& 	$\times$ 		&	$\checkmark$	&	$\checkmark$	& 	$\times$		\\ \hline
           M-CAS 		&	$-$				& 	$\checkmark$ &	$\checkmark$ &	$\checkmark$ 	&	$\checkmark$	& 	$\checkmark$ 	& 	$\checkmark$ 	& 	$\times$		\\ \hline
           SAS 		&	$\checkmark$	& 	$\checkmark$ &	$\checkmark$ & 	$\checkmark$ 	&	$-$ 			&	$\checkmark$	& 	$\checkmark$ 	& 	$-$				\\ \hline
          {\bf Proposed (SubLock)} 		&	{\bf $\checkmark$} 	&	{\bf $\checkmark$} & {\bf $\checkmark$} & {\bf$\checkmark$}	& 	{\bf $\checkmark$ }	&	{\bf $\checkmark$} 	&	{\bf $\checkmark$} 	& {\bf $\checkmark$ }	\\ \hline
         \end{tabular}}
       \end{table*}

    It is also reported in the literature that SAS and SFLL provide low output corruptibility. Therefore, two methods called CORruption adaptable logic locking (CORALL) \cite{9070188} and DisORC \cite{9214869} are reported that increased the output corruptibility in comparison to SAS and SFLL techniques. In addition to SAT attacks, these methods also increase the security against structural or netlist analysis attacks. Besides these methods, the generalized Anti-SAT (G-Anti-SAT) \cite{zhou2021generalized} and SKG-Lock+ \cite{nguyen2022skg} are also reported, which significantly increase output corruption without compromising the security of logic locking against SAT and other related attacks. A ChaoLock method is reported in \cite{kamali2021chaolock} that uses asymmetric chaotic Boolean gates and dummy inputs to increase the security against SAT attack.
   
   However, the G-Anti-SAT is susceptible to a vulnerability assessment tool called Valkyrie \cite{limaye2022valkyrie}, SigAttack \cite{shen2019sigattack} and generalized \cite{patnaik2022hide} attacks and SKG-Lock+ and ChaoLock incur around 10\% area overhead in most of the ISCAS-85 benchmarks even with low size key. In addition to these techniques, some other SAT-resistant logic locking methods are proposed. In \cite{shamsi2017cyclic}, the authors have added the dummy cycles in the locked design to significantly increase the attacker's efforts while deciphering the correct key using SAT attack. Since cyclic logic locking failed against the CycSAT attack \cite{zhou2017cycsat}, the unreachable states are added in \cite{rezaei2019cycsat} to improve the security of cyclic logic locking against the CycSAT attack. Recently, an extension of cyclic logic locking called LoopLock 2.0 \cite{9335625} and LoopLock 3.0 \cite{10473877} are proposed to enhance the security against the CycSAT and SAT attacks simultaneously. However, LoopLock 3.0 provides better security than LoopLock 2.0. 
 
  All the previously proposed techniques primarily focus on increasing the number of SAT iterations rather than preventing them at the cost of significant overhead. This is because the traditional SAT attacks work in such a manner that it requires exponential time/iterations to break them. However, the analysis of the SAT attack presented in \cite{article23} shows that the SAT attack can defeat any logic locking method in linear time by using a different relation between keys using DIPs and their respective oracle responses. Hence, it is highly desirable to develop a method that can prevent the SAT attack instead of just enhancing security against SAT attacks. However, the input-dependent key-based logic locking (IDKLL) method called GateLock is reported in \cite{10375951} that attempts to prevent the SAT attacks. Since this method uses IDKLL-based gates, thus it may be vulnerable to structural analysis attacks. Therefore, this paper proposes a novel sub-circuit replacement-based IDKLL method called SubLock to effectively prevent the SAT attack and its variants with less overhead and provide robust IP protection. Due to using sub-circuit replacement, the proposed method also provides effective security against structural analysis-based attacks compared to GateLock. The comparison of the efficacy of proposed and existing methods for mitigating different attacks is presented in Table \ref{Tabcomp}. The next section discuss the concept of IDKLL in details and its use in logic locking followed by the proposed SubLock method.

\section{IDKLL: INPUT DEPENDENT KEY-BASED LOGIC LOCKING} \label{Prop-1}
  In the previous logic locking techniques, the locked design provides the correct output for all the input patterns when a valid key sequence is applied. A key sequence is considered correct or valid if it gives the correct output for all the input patterns; otherwise, it is incorrect. Therefore, the designer stores only that valid key sequence in a tamper-proof memory to activate or unlock the design. The SAT-based attacks eliminate the incorrect key sequences and identify the correct key sequence by iteratively applying DIPs \cite{yasin2016sarlock}, \cite{RATHOR2020Anti-SAT}. Since DIP is an input pattern that provides different outputs for two key sequences, SAT attack eliminates a key sequence that provides incorrect output. This process repeats until all the wrong keys are eliminated. The computation time of SAT attacks mainly depends on the number of iterations or DIPs required to eliminate the incorrect keys. Though the existing logic locking techniques increase SAT attack computation time, they fail to prevent it completely. Therefore, this paper introduces a novel idea of logic locking called input dependent key-based logic locking (IDKLL) that effectively prevents SAT-based attacks.
 
 \subsection{Concept of IDKLL} 
   The proposed idea for preventing the SAT attack is to lock a design utilizing several key sequences (rather than a single key sequence) as the valid key so that the locked design only gives correct output for all input patterns when their respective valid key sequence is applied. In other words, each set of input patterns has a separate correct key sequence in the locked design. As a result, the inputs determine the correct key sequence for unlocking the design. It implies that a key sequence can only unlock the design for a single set of input patterns. A different key sequence must be used to unlock the functionality for the second set of input patterns. Consequently, other key sequences could be used to unlock the design functionality for the third, fourth, and other sets of inputs. Consider a circuit with $n$ primary inputs and $k$ key-inputs supposed to be locked. Suppose we employ $m$ key sequences out of a total of $2^k$ to unlock the design. In that case, the $2^n$ input patterns must be separated into $m$ sets so that the locked design gives accurate output for each input set when its corresponding valid key sequence (KS) is applied. If we choose key sequences $KS_1$, $KS_2$,..., $KS_m$ sequences as valid for input patterns sets $XS_1$, $XS_2$,..., $XS_m$ respectively, then the locked design produces correct output for $XS_1$, $XS_2$,..., $XS_m$ only when $KS_1$, $KS_2$,..., $KS_m$ key sequences are applied respectively. Moreover, there must not exist any key sequence that can unlock the design or provide correct output for all input patterns. The correct outputs for all the input patterns can only be generated when their corresponding key sequences are applied.
 
  %However, we cannot lock the circuit with input dependent key using the key-gates insertion or replacements based logic locking. Because these approaches only use a single key sequence as correct to unlock the design for all the input patterns, or the design provides correct output for all patterns only with a single correct key sequence. Therefore,
  
  \subsection{Locking a Design using IDKLL}
\begin{figure}[h!]    \centering
     \includegraphics[width=0.75\columnwidth]{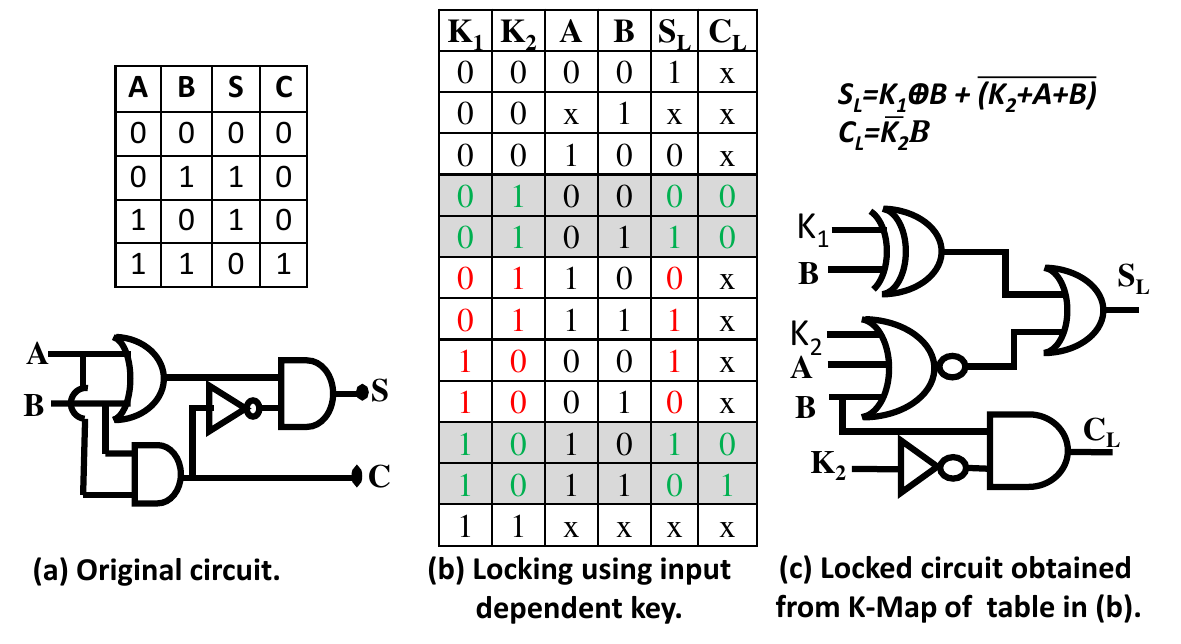}
      \caption{Locking a half adder circuit using input dependent key-based logic locking. The locked circuit provides correct output for the input patterns $\{``00", ``01"\}$ and $\{``10", `11"\}$ only when the two key sequences $K_1 K_2 = ``01"$ and $``10"$ are applied respectively as a valid key.}
    \label{fig1}
  \end{figure}

   To use IDKLL to lock the circuit, we make the original circuit's functionality dependent on key inputs such that circuit only produces correct output for all input patterns when their respective valid key sequence is applied. For example, consider a half-adder circuit as shown in Figure \ref{fig1}(a). We lock this half-adder circuit using input dependent key-based logic locking, as shown in Figure \ref{fig1}(c). To lock the functionality of this circuit, we divide the input patterns into two sets, $i.e.$, $XS_1=\{``00", ``01"\}$ and $XS_2=\{``10", ``11"\}$. The original circuit is locked by making the functionalities of its original outputs, $i.e.$, sum (S) and/or carry (C), dependent on two key-inputs $K_1$ and $K_2$ according to the truth table as shown in Figure \ref{fig1}(b). Here, we use $m=2$ key sequences $KS_1 = ``01"$ and  $KS_2 = ``10"$ as a valid key for the input sets $XS_1$ and $XS_2$ respectively. The locked functionality of circuit is represented with $S_L$ and $C_L$, the correct output would be obtained for the input sets $XS_1=\{``00", ``01"\}$ and $XS_2=\{``10", ``11"\}$ only when the key sequences $KS_1 = ``01"$ and  $KS_2 = ``10"$ will be applied respectively. It is also ensured that $KS_1$ and $KS_2$ only provide correct output for the specified set of inputs and must provide complementary or incorrect output for the other set of inputs. Otherwise, it may be possible that a single key sequence provides the correct output for all inputs. Therefore, we keep $S_L$ as complementary of $S$ ($i.e.$, $S_L = !S$)  in this example for the inputs \{``0110", ``0111", ``1000", ``1001"\}. 
 
   Since here we mainly lock the functionality of the sum (S) logic of the half adder circuit, we keep the $C_L$ don't care (x) for all the other input patterns except \{``0100", ``0101", ``1010", ``1011"\}. It is observed that the overhead for implementing IDKLL can be reduced when a design is locked by keeping the output value don't care for the invalid key sequences. Therefore, we also lock the circuit functionality ($S_L$ and $C_L$) by keeping them don't care (x) for the remaining key-values ($i.e.$, ``00" and ``11"). But due to optimization, the expression (obtained from the K-map) of $S_L$ does not depend on all the inputs ($i.e$, A and B). Thus, to ensure the dependency of $S_L$ on both inputs (A and B), we keep $S_L$ `1'/`0' instead of don't care (x) for a few patterns, $i.e.$, ``0000" and ``0010". However, a designer can also select different patterns to do the same. If the designer is not worried about the dependency, then he/she can keep don't care for the remaining key values. 
   
\begin{figure}[h!]    \centering
     \includegraphics[width=0.6\columnwidth]{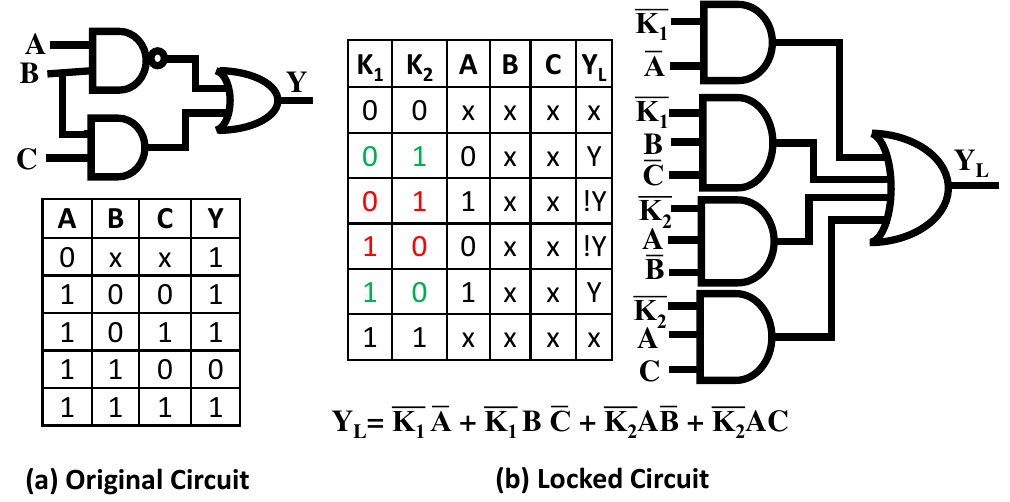}
     \caption{An example of input dependent key-based logic locking, where the output bits are kept don't cares (x) for all the remaining incorrect key values, $i.e.$ ``00" and  ``11". The locked circuit provides correct output for the sets of input patterns \{``000", ``001", ``010", ``011"\} and \{``100", ``101", ``110", ``111"\} only when ``01" and ``10" key sequences are applied respectively.}
      \label{fig2}
   \end{figure}

   In case the locked functionality already depends on all inputs while retaining, don't care (x) for incorrect key-values. Then, we should keep don't cares (x) to produce an optimized locked design as illustrated in Figure \ref{fig2}. In this figure, the original circuit is locked using two key inputs $K_1$ and $K_2$, where $m=2$ key sequences, $i.e.$, \{``01", ``10"\} out of four key sequences, $i.e.$, \{``00", ``01", ``10", ``11"\} are selected to provide correct output for the sets of input patterns, $i.e.$, \{``000", ``001", ``010", ``011"\} and \{``100", ``101", ``110", ``111"\} respectively. It is clear from this figure that the Boolean expression of the locked circuit already depends on all the inputs while keeping $Y_L$ don't care for the reaming key sequences, $i.e.$ ``00" and ``11". We also lock this circuit by keeping output bits `0'/`1' for the incorrect key sequences with different input combinations. Finally, we discovered that compared to all other locked circuits, the circuit locked by preserving the output value don't care for the rest of the key sequences always requires fewer number literals. However, keeping the wrong output (!Y) instead of don't care (x) for the remaining key sequences results in high output corruption. As a result, one significant advantage of our method is that it allows the designer to achieve desired output corruption, which may not be possible with existing logic locking solutions.
  
  \subsection{Use of LUT based Tamper-Proof Memory for IDKLL}  
\begin{figure}[h!]    \centering
      \includegraphics[width=0.6\columnwidth]{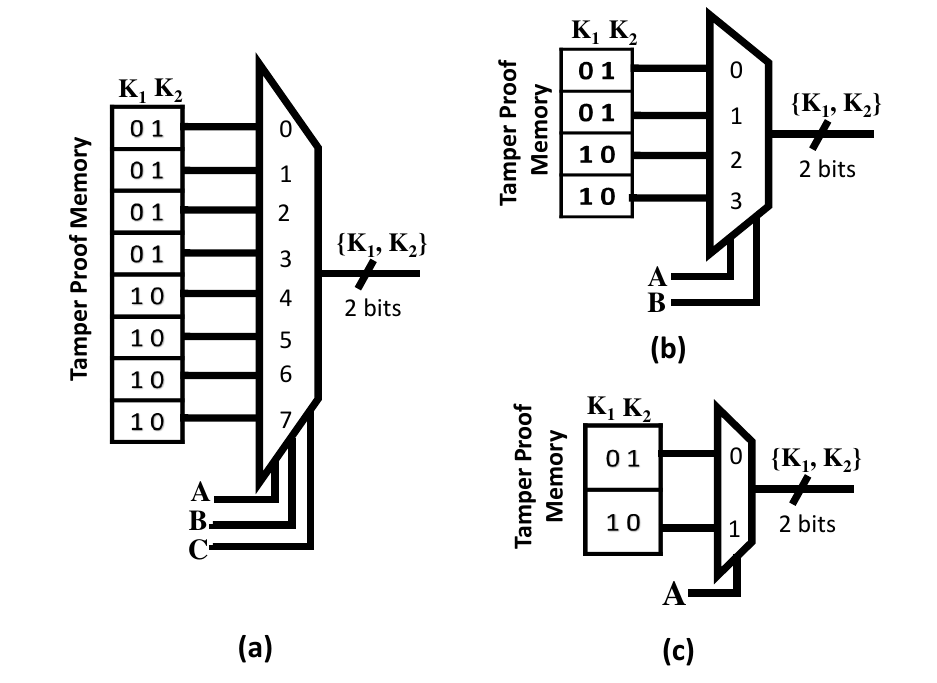}
       \caption{The structure of LUT based memory for extracting the correct key sequence for the locked circuit shown as in (a) Figure \ref{fig2} (b) Figure \ref{fig1} (c) Optimized structure of LUT based memory.}
      \label{fig3}
    \end{figure}

   In the prior logic locking works, a single key sequence was employed as a valid key to unlock the design for all the input patterns. As a result, the designer stores a single key sequence in the tamper-proof memory. However, the proposed logic locking method unlocks the design by using numerous key sequences as the valid key. Therefore, our method keeps all of these correct key sequences in a tamper-evident memory to unlock the design for all input patterns. The different key sequences are valid for different input patterns. Therefore, we employ LUT-based tamper-proof memory to extract the correct key sequence for the applied input pattern. Figure \ref{fig3}(a) and (b) presents the use of LUT based memory for the locked circuits presented in Figure \ref{fig2} and Figure \ref{fig1}, respectively.
  
   It is clear from Figure \ref{fig3}(a) that for the set of input patterns \{``000", ``001", ``010", ``011"\} the correct key sequence, $i.e.$, ``01" can be easily fetched from the tamper-proof memory by applying any of these patterns at the selection line of the multiplexer. In the same manner, Figure \ref{fig3}(b) shows the extraction of the right key sequence for the applied input. However, it can be realized from Figure \ref{fig3}(a) and \ref{fig3}(b) that the key sequences, $i.e.$ ``01" and ``10", are repeated numerous times in memory, which results in increased design cost. Therefore, to save the required memory, we should store unique, valid key sequences as shown in Figure \ref{fig3}(c). Moreover, in Figure \ref{fig3}(a) and\ref{fig3}(b), the value of the correct key for different patterns only changes (from ``01" to ``10") with input $A$ and it is independent of the other inputs, $i.e.$ $B$, $C$. Therefore, we remove these two inputs for optimized implementation of LUT based tamper-proof memory to store correct key sequences ``01" and ``10" as shown in Figure \ref{fig3}(c). 
   
  However, one thing can also be observed from the above that using LUT based tamper-proof memory requires more overhead compared to the conventional normal tamper-proof memory. This is because the LUT based memory stores multiple key sequences valid for different input sets. Whereas the conventional normal memory stores only a single key sequence that is valid for all input patterns. Further, LUT based memory uses an additional multiplexer unit to fetch the respective valid key for different sets of inputs. Due to these reasons, the LUT based tamper-proof memory requires more overhead than the conventional normal tamper-proof memory. However, analysis of this is out of scope in this paper. This paper uses LUT based memory as one solution to store multiple key sequences and fetch respective valid key sequences on applying an input. One can also explore any other cost-effective solution to achieve the same. The next section presents the generalized structure of IDKLL and its integration with LUT based memory.

  \subsection{Complete Generalized Structure of IDKLL after Integrating LUT based Memory}  
\begin{figure}[h!]    \centering
      \includegraphics[width=0.6\columnwidth]{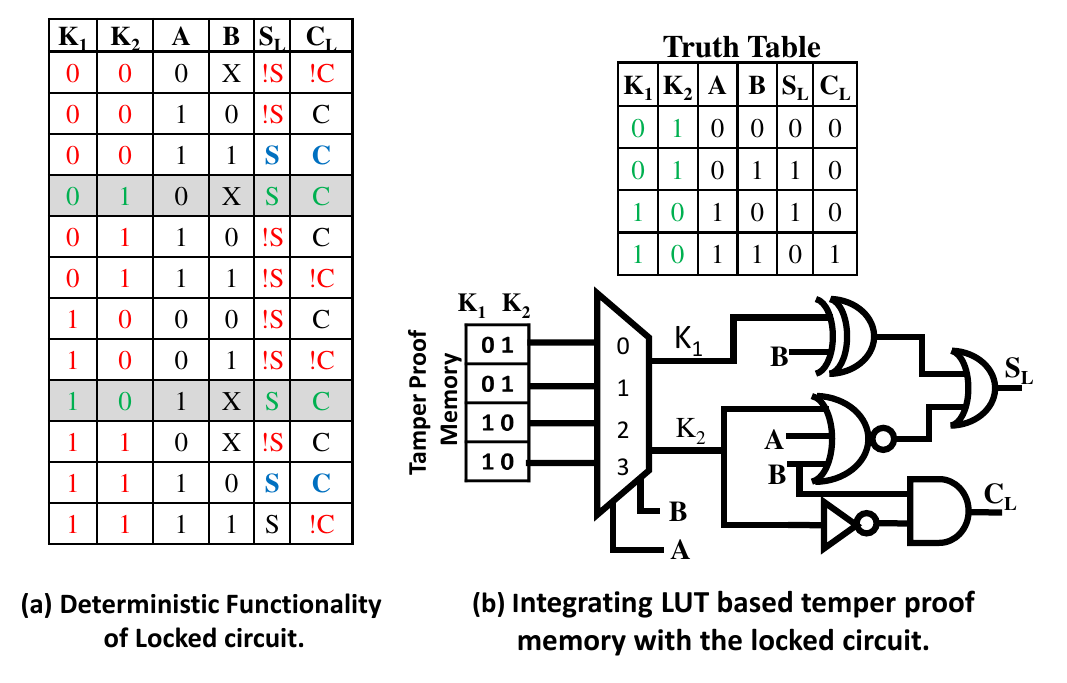}
       \caption{Example for integrating LUT based tamper-proof memory with the locked circuit to achieve correct functionality.}
     \label{fig3_1}
   \end{figure}

   The truth tables of the above-locked circuit, as shown in Figure \ref{fig1}(b) and Figure \ref{fig2}(b), exhibit don't care values for some outputs. However, the boolean expression or circuit obtained after solving the K-Map will always produce deterministic outputs. Therefore, we also present the truth table in Figure \ref{fig3_1}(a), which exhibits deterministic output for every input of the locked circuit/expression as shown in Figure \ref{fig1}(c). We observed from this table that the locked circuit also provides correct output for the inputs ``11" and ``10" on applying key sequences``00" and ``11" respectively. Though these key sequences cannot produce the correct output for all patterns, they may be used in combination with other valid key sequences to achieve the correct output. For example, utilizing ``00" and ``11" key sequences (instead of ``10") along with ``01" (i.e., key set \{``00", ``01", ``11"\}) can also achieve correct output for all patterns i..e, ``11", ``10" and \{``00", ``01"\} respectively. Hence, it cannot be ensured that only a single valid key set exists that provide correct output for a locked circuit. However, one can easily ensure that only a single valid set can lock/unlock the design at the cost of increased overhead. To do this, we can specify the complementary (i.e., wrong output) output instead of don't care (x) for all the remaining/unused key sequences (which are not part of valid key sequences). In this case, only valid key sequences will only provide correct output for the input patterns, and all the unused/remaining (invalid) key sequences will guaranteed provide incorrect output for the input patterns. However, the locked design requires a large overhead if we mention complementary/wrong output for all unused key sequences. Therefore, the proposed method use don't care (x) for the remaining key sequences to obtain the optimized design and ensure that no single key sequence exists that provides correct output for all input patterns by manually verifying the truth table of synthesized locked sub-circuits.

    Since none of the single key sequences generates the correct output, the attacker cannot apply SAT attack, or SAT attack will either return a single key that will not make it successful. In addition to the above, we cannot ensure in the proposed method that a returned key sequence does not belong to the valid key set because SAT attack always eliminates wrong key sequences and retain the key sequence, which provides the correct output for at least one pattern. However, if a key sequence provides correct output for a few patterns, then still SAT attack cannot be applied/worked on the proposed method even if returned key sequences provide correct output for a few patterns or belong to a valid key set. This is because, in our method, all key sequences are incorrect individually; SAT attack eliminates all key sequences and returns a single key sequence. Though in our method, the returned key sequence may belong to a set of correct key sequences or a set of incorrect/unused key sequences, it can never provide correct output for all patterns, which will make SAT unsuccessful. Therefore, in our method, we do not consider that the key returned by SAT attack must not belong to the valid key set.

   Finally, we combine the locked design with the LUT based tamper-proof memory to obtain the correct output. The integration of the locked circuit(shown in Figure \ref{fig1}(c)) with its LUT based tamper-proof memory is given in Figure \ref{fig3_1}. It is clear from this that correct output can be easily obtained for all patterns, as shown in Truth Table. 
 
\begin{figure}[h!]    \centering
      \includegraphics[width=0.5\columnwidth]{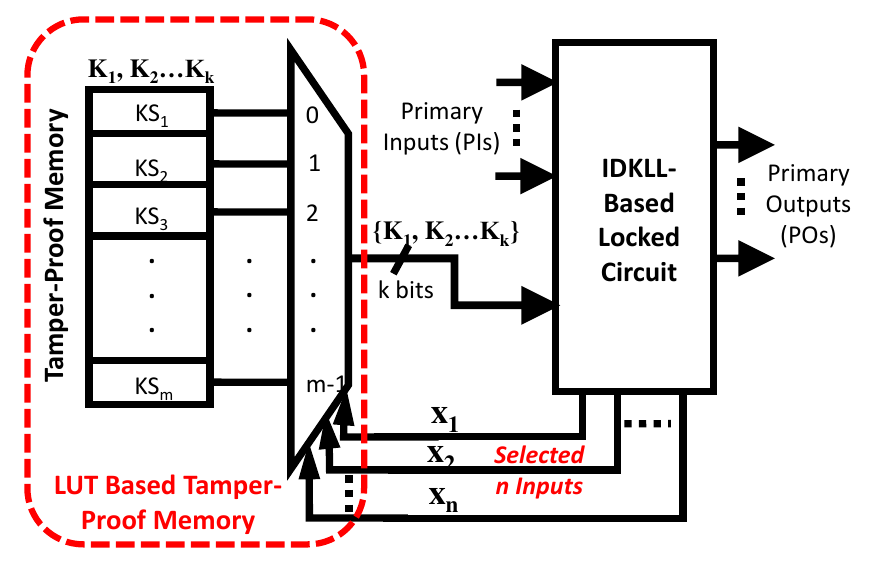}
      \caption{Generalized structure and integration of LUT based tamper-proof memory with the design locked using IDKLL.}
     \label{fig4}
   \end{figure}

  Furthermore, Figure \ref{fig4} shows the generalized structure of LUT based tamper-proof memory and its integration for generating the input dependent key (or storing valid key sequences) for unlocking the design locked using the proposed IDKLL. Only the key sequences that produce correct output are considered valid and stored in this memory. In the proposed IDKLL, all the key sequences are individually invalid, and no key sequence exists that can provide the correct output for all the inputs. Therefore, the attacker will fail to determine the valid key sequences by applying DIPs in SAT-based attacks. This is because the application of a DIP produces different outputs for two valid key sequences. The SAT solver eliminates one correct key sequence, which provides incorrect output for that DIP. Similarly, the second DIP will eliminate the second key sequence. Finally, the SAT solver either eliminates all the key sequences or returns a single valid/invalid key sequence. The returned key sequence would be insufficient to obtain the correct output for all the inputs because the locked circuit cannot produce the correct output for all the inputs until all their respective correct key sequences are applied. 
    
\begin{figure}[h!]    \centering
      \includegraphics[width=0.5\columnwidth]{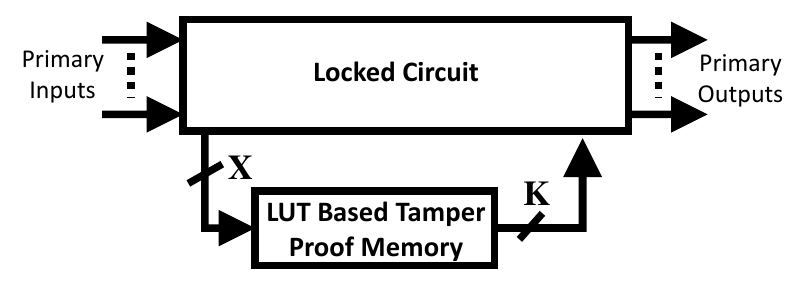}
        \caption{Block level implementation of proposed input dependent key-based logic locking approach.}
       \label{fig5}
    \end{figure}

 The implementation of the proposed IDKLL and LUT-based tamper-proof memory for storing the valid key sequences for the small circuit is explained above. The concept of LUT based tamper-proof memory can also be easily represented at the block level, as shown in Figure \ref{fig5}.
       
  However, locking a small circuit using the proposed IDKLL is explained above. Locking a large design that exhibits many inputs/outputs using the above procedure (where we also make the output dependent on the key inputs) would be very complex and challenging. Further, it may also require significant overhead, which may be unaffordable sometimes. Therefore, we also propose a low-cost approach for locking a design using input dependent key-based logic locking in the next section.

%\section{LOW-COST IMPLEMENTATION of IDKLL and SECURITY ANALYSIS} \label{Prop-2}
%   This section first presents the low-cost approach for implementing the input dependent key-based logic locking. Further, we present the quantitative security analysis of the proposed IDKLL technique.
 
%\subsection{SUB-Lock: Sub-Circuit Replacement based IDKLL} \label{Prop_light}
\section{SUB-Lock: Sub-Circuit Replacement based IDKLL} \label{Prop_light}
  The locking design by considering the whole functionality (discussed in Section \ref{Prop-1}) is very challenging and may require unaffordable design overhead. However, we can insert an additional locked circuit similar to Anti-SAT to lock the design functionality. The additional insertion may also incur a large overhead. Therefore, we propose an approach that locks the design by replacing the original sub-circuitry of the design with the corresponding new circuit that is locked using input dependent key-based logic locking. In the proposed approach, we select several sub-circuits in a design and then the selected circuits are replaced with their corresponding locked circuits, which are locked using proposed logic locking as discussed in Section \ref{Prop-1}. 
      
\begin{figure}[h!]    \centering
      \includegraphics[width=0.9\columnwidth]{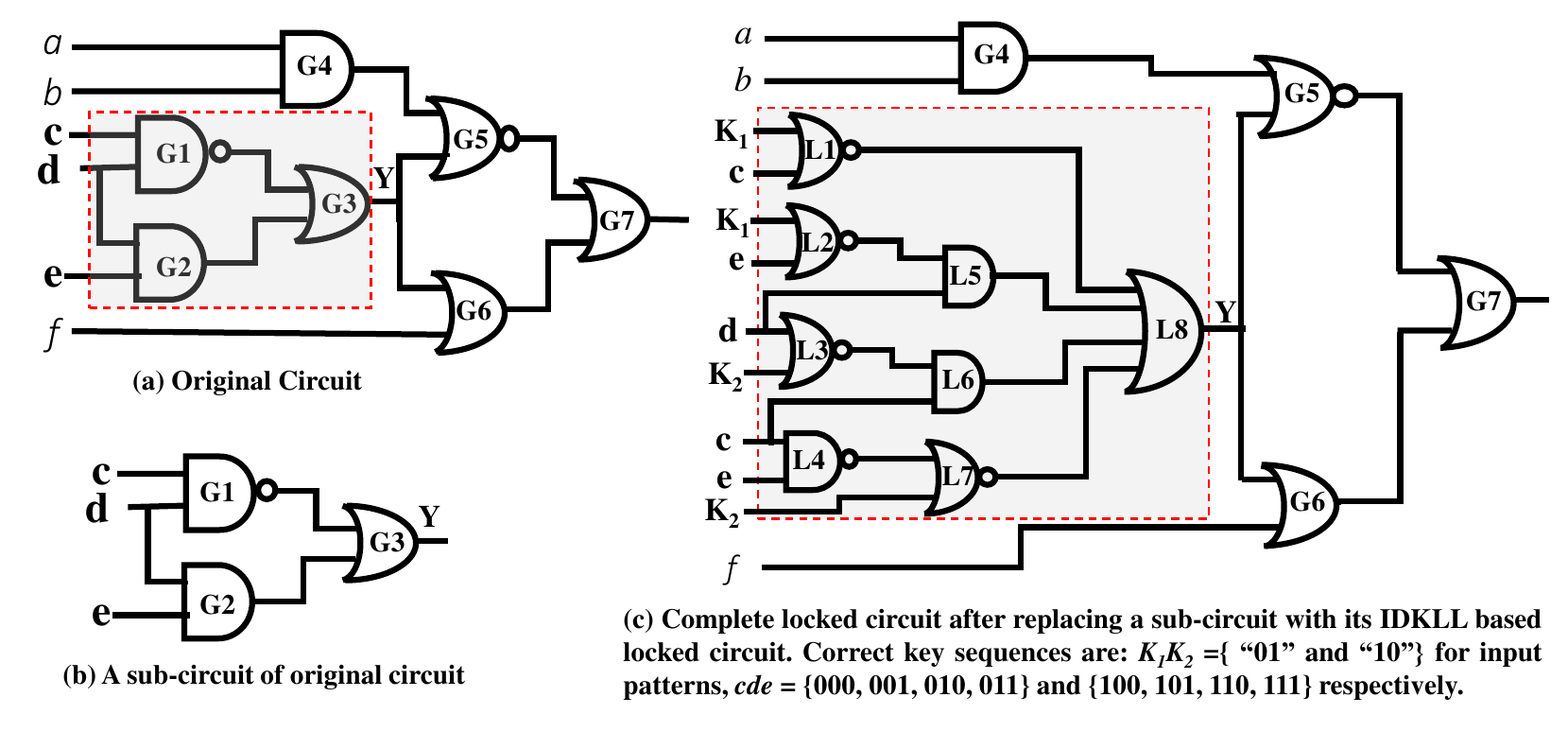}
       \caption{The example of locking a design by replacing a small sub circuit with their corresponding IDKLL based locked circuit.}
       \label{fig7}
   \end{figure}

    The example of locking a design by replacing its sub-circuit is shown in Figure \ref{fig7}. Here, an example circuit, as shown in Figure \ref{fig7}(a), is considered for locking by replacing its sub-circuit ($Y = \overline{cd} + de$)  with the corresponding circuit, which is locked using input dependent key-based logic locking. To achieve this, we first construct the lock version of the selected sub-circuit as shown in Figure \ref{fig7}(b) using proposed input dependent key-based logic locking. Finally, the original circuit is locked by replacing the selected sub-circuit ($Y = \overline{cd} + de$) with its locked version as shown in Figure \ref{fig7}(c)). This figure presents the complete locked design, where the locked version of the originally selected sub-circuit is mentioned with the dotted box. In this locked circuit, if we analyzed the overhead in terms of gate count, then only three additional gates are required due to replacing the existing sub-circuitry. Hence, locking a design by replacing multiple sub-circuits with their corresponding locked circuits can significantly reduce the design overhead compared to the insertion-based logic locking technique, $i.e.$, Anti-SAT \cite{xie2018anti}.

  The selection of sub-circuits for locking the design can be random or judicious. Though random selection can introduce some non-determinism while locking the design, it may fail to provide high output corruption. On the other hand, a judicious replacement can provide high output corruption. For example, selecting the sub-circuits whose output exhibits the highest fault impact point \cite{rajendran2015fault} for replacement can significantly increase output corruption. The designer can combine both the above approaches to achieve high security. It can also be observed in proposed logic locking that no additional special structure or key-gates ($i.e.$, XOR/XNOR, MUX etc.) are inserted \cite{juretus2016reduced}, \cite{rajendran2015fault}. Hence, it would be very difficult to identify the replaced/locked circuitry in the design exactly.
  Moreover, the designer can also lock the design by inserting the additional locked circuits and by replacing the existing sub-circuits with their locked circuits, as reported in \cite{RATHOR2020Anti-SAT}. In this case, it is very difficult for the attacker to distinguish which part of the circuit is additionally inserted or which part is an existing part. The quantitative security analysis of the proposed logic locking is given in the next section. 

\section{Security Analysis of Proposed Technique}     
%\subsection{Security Analysis of Proposed Technique}  
  It is clear from the above that the proposed input dependent key-based logic locking technique completely prevents SAT-based attacks. The attacker cannot apply the SAT-based attack to decipher the correct key from the circuit locked using the proposed input dependent key-based logic locking. Hence, the attacker has to employ only a brute force attack to extract the key. Therefore, in this subsection, we also quantitatively analyze and compare the security of the proposed technique against the brute force attack.
  Let us consider the above discussion as shown in Section \ref{Prop-1}, where a circuit with $n$ primary inputs is locked by inserting $k$ key-inputs. If this design is locked using standard logic locking techniques, the attacker has to apply all key sequences or brute force attempts to unlock the design \cite{rajendran2015fault}, \cite{yasin2016improving} \cite{karmakar2017new}. If the total key sequences or brute force attempts for $k$ bit-length key are denoted with $l$, then $l$ can be represented as
    \begin{equation} \label{eq1}
    	l = 2^k
    \end{equation}
  
  Although, the SAT-based attacks can break these standard logic locking techniques within a few attempts \cite{subramanyan2015evaluating}. The use of SAT-resistant logic locking techniques such as Anti-SAT \cite{xie2016mitigating}, \cite{xie2018anti} forces the attacker to apply at least  $\lambda$ attempts/iterations to extract the correct key sequence. Here, the value of  $\lambda$  for $k$ bits key is represented with Eq. (\ref{eq2}) \cite{xiao2016hardware}.
  
    \begin{equation} \label{eq2}
      	\lambda = 2^{k/2}
    \end{equation}
  
  On the other hand, if we use $m$ key key-sequences out of $l$ to lock the entire design functionality using the proposed technique, then to know the correct key sequences and their order, the attacker has first to know the output corresponding to each key sequence. Next, he/she has to identify correct combinations of key sequences as well as their order of application as a secret key, such that it provides correct output for all input patterns. The number of attempts required to know the output of each key sequence would be equivalent to the brute force attempts, $i.e$, $l =2^k$ as shown in Eq. \ref{eq1}. Though we utilize $m$ key sequences as a correct key out of $l$, the attacker would not be aware of how many key sequences, which combination, and which order ($i.e.$ which key sequence is used for which input pattern) they have used to lock design. Therefore, the attacker has to apply all possible permuted combinations of $l$ key sequences to determine a correct combination and order of application of key sequences. If the total permuted combinations for $l$ key sequences are denoted with $PC$, then $PC$ can be represented as follows. 
     
  \begin{equation} \label{eq3}
     PC = \Mycomb[l]{1}.l! + \Mycomb[l]{2} .2! + \Mycomb[l]{3} .3! + ..... + \Mycomb[l]{l} .l!
  \end{equation}   
 
 The above equation can also be represented in the form of permutation as follows
  \begin{equation} \label{eq4}
     PC = \Myperm[l]{1} + \Myperm[l]{2} + \Myperm[l]{3} + ..... + \Myperm[l]{l}
  \end{equation}   
  
 Since,
 \[ \Myperm[l]{r} = \frac{l!}{(l-r)!} \]
 
 Thus, we can also write the Eq. \ref{eq4} as follows.
 
  \begin{equation} \label{eq5}
      PC = \frac{l!}{(l-1)!} + \frac{l!}{(l-2)!} + \frac{l!}{(l-3)!} + ..... + \frac{l!}{(l-l)!}
   \end{equation}  
   
   \begin{equation} \label{eq6}
     PC = l! \left(\frac{1}{(0)!} + \frac{1}{(1)!} + \frac{1}{(2)!} + ..... + \frac{1}{(l-1)!}\right)
   \end{equation} 
   
   \begin{equation} \label{eq7}
        PC = l! \sum_{r=1}^{l-1} \frac{1}{(r)!}
  \end{equation}    
   
   However, the above equation can also be represented in other forms; this is out of scope in this paper. Thus, we compute the total attempts required by the attacker to decipher the correct key sequences by adding the attempts of finding the output, $i.e.$, $l$ (Eq. (\ref{eq1})) with the total permuted combinations (Eq. (\ref{eq7})) of key sequences as follows.
    \begin{equation} \label{eq8}
        	Total\_Attempts = l + PC
    \end{equation}
  
  Finally, the total number of brute force attempts required by the attacker to extract correct key sequences in the proposed technique can be computed by placing the value of $l$ and $PC$ from Eq. (\ref{eq1}) and Eq. (\ref{eq7}) into the Eq. (\ref{eq7}) as follows
    \begin{equation} \label{eq9}
     Total\_Attempts = 2^k  + (2^k)!\sum_{r=1}^{2^k-1} \frac{1}{(r)!}
    \end{equation}
    
  Now, it can be observed from Eq. (\ref{eq1}) and Eq. (\ref{eq2}) that existing logic locking techniques are only able to maintain $2^k$ brute force attempts without preventing SAT attack. They can provide security against SAT attacks while maintaining only $2^{k/2}$ attempts/iterations. On the other hand, the proposed technique significantly increases the number of brute force attempts for the attacker to decipher the correct key over the existing techniques while preventing SAT attack, as shown in Eq. (\ref{eq9}). This equation basically presents the attack complexity for the entire locked circuit. The attacker may also attempt to identify the locked sub-circuits/sub-cones and break them individually. However, it will require knowing the original functionality of the replaced sub-circuit. But due to replacement, it will be approximately impossible for the attacker to know the original functionality of the replaced sub-circuit. Further, due to the use of multiple key sequences as valid for each sub-circuit, the attacker cannot apply SAT attack even on individual sub-circuits. Therefore, the attacker must apply brute force similar to Eq. (\ref{eq9}), which will require exponential complexity.
  
  Due to the replacement, if the attacker also tries to break the multiple sub-circuits simultaneously, he/she also cannot solve them simultaneously. Further, the output of one locked sub-circuit may also interfere with the output of another locked sub-circuit. Due to this and using multiple key sequences as the valid key, the attacker cannot even apply sensitization and any other attack to break the proposed IDKLL method. Hence, an attacker will always require exponential complexity (or brute force) to break the proposed method in every case. In the proposed method, the number of increased brute force can be computed by subtracting the Eq. (\ref{eq1}) from Eq. (\ref{eq9}) as shown below.
  
  \begin{equation} \label{eq10}
       Increased\_Attempts = (2^k)!\sum_{r=1}^{2^k-1} \frac{1}{(r)!}
  \end{equation}

It is clear from the above equation that the proposed input dependent key-based logic locking technique prevents SAT-based attacks and significantly increases the required number of brute force attempts over all the existing techniques. We also justify this claim with the experimental evaluation of the proposed technique, as presented in the next section.

\section{Experimental Results and Analysis} \label{exp}
  This section first presents the experimental setup, followed by simulation results and comparative analysis.

 \subsection{Experimental Setup} 
   We locked the ISCAS/ITC benchmark circuits by randomly replacing the 3-input and 4-input sub-circuities with their corresponding locked circuits which are locked using the proposed IDKLL method as discussed in Section \ref{Prop_light}. To analyze the effectiveness of the proposed technique, the varying number of keys (i.e., 16, 32, 64, 128) are embedded in the benchmarks, and different locked circuits are generated, namely Prop\_16K, Prop\_32K, Prop\_64K and Prop\_128K. The security of the proposed technique against SAT-based attacks is evaluated using the SAT attack tool as reported in \cite{subramanyan2015evaluating}. Finally, to analyze the implementation overhead, the locked circuits are synthesized using Cadence RTL compiler with Nangate Open Cell Library 45nm \cite{nagate11}, and different design metrics (area, power, delay) are extracted. We compared our approach with the existing methods. 
     
\subsection{SAT Attack Results and Analysis}
\begin{figure}[h!]  \centering
%      \subfigure[SAT attack results on c7552 locked circuit with 16-bit key]{\includegraphics[width = 0.7\columnwidth]{figures/fig9_a.png}}
%      \subfigure[SAT attack results on c1908 locked circuit with 16-bit key]{ \includegraphics[width = 0.7\columnwidth]{figures/fig9_b.png}}
%      \subfigure[Validation of key resturned by SAT attack for c1908 locked circuit]{ \includegraphics[width = 0.7\columnwidth]{figures/fig9_c.png}}
      
    \subfigure[SAT attack results on c7552 locked circuit with 16-bit key]{\includegraphics[width = 0.7\columnwidth]{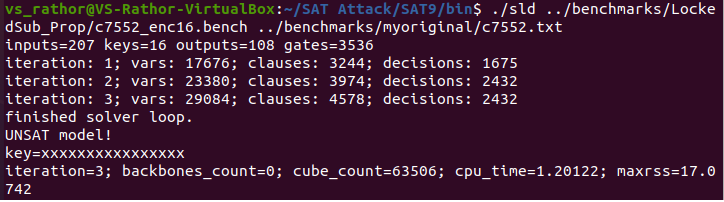}}
    \subfigure[SAT attack results on c1908 locked circuit with 16-bit key]{ \includegraphics[width = 0.7\columnwidth]{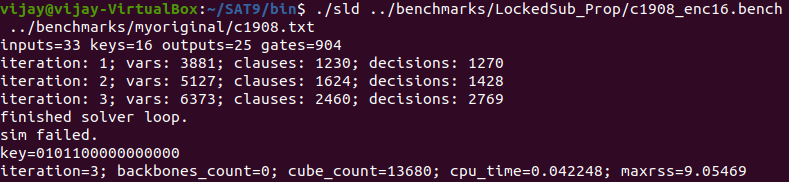}}
    \subfigure[Validation of key resturned by SAT attack for c1908 locked circuit]{ \includegraphics[width = 0.7\columnwidth]{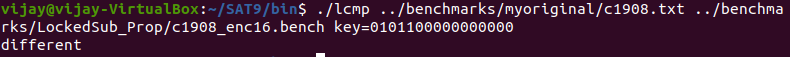}}
     \caption{Sample snapshots of the demonstration of SAT attack on the benchmarks locked using proposed sub-circuit replacement based IDKLL.}
       \label{fig9}
 \end{figure}

To validate the effectiveness of the proposed method against SAT attacks, the locked versions of the selected sub-circuits are also individually verified for functionality and security against SAT attacks. Afterwards, we locked the benchmark circuits by randomly replacing their sub-circuits with the corresponding IDKLL based locked versions. We have embedded the different number of keys (16, 32, 64, 128) in each benchmark while locking them. Finally, we evaluate the effectiveness of the proposed sub-circuit replacement based IDKLL method against SAT attack by applying SAT attack on the above-locked benchmarks using SAT solver tool \cite{subramanyan2015evaluating}. As a result of the SAT solver, we found that the SAT attack cannot extract the correct key from the individual locked sub-circuits as well as from the locked benchmarks circuits. It either provides the ``UNSAT Model" or returns a key that has been proven wrong on its verification. The screen-shots of the demonstration of SAT attack on the c7552 and c1908 benchmark circuits locked by inserting 16 keys are shown in Figure \ref{fig9}.

 SAT attack completely failed to identify the correct key even with applying the Partial-Break algorithm along with fault analysis \cite{rajendran2015fault}. It can be observed from this figure that the application of SAT attack on the c7552 locked benchmark provides the ``UNSAT Model", which means it is unable to determine the correct key as shown in Figure \ref{fig9}(a). Although SAT attack returns a key for the c1908 locked benchmark (Figure \ref{fig9}(b)), the returned key is declared wrong or incorrect after verification using SAT solver program as shown in Figure \ref{fig9}(c). Apart from c7552 and c1908, we have also applied the SAT attack on other locked ISCAS and ITC benchmark circuits. We observed that SAT attack has completely failed to identify the correct key in any locked benchmark circuit that is locked using the proposed IDKLL. The SAT attacks simulation results for ISCAS benchmarks locked with 16 and 32 keys are shown in Table \ref{Tab4}.
  
   \begin{table}[h!]
      \centering
      \caption{Analysis of Proposed Technique using SAT Attack Tool}
      {\small
       \label{Tab4}
       \begin{tabular}{|c|c|c|c|c|c|c|}   \hline
         \multirow{2}{*}{Circuits} &  \multicolumn{3}{c|}{Prop\_16K} & \multicolumn{3}{c|}{Prop\_32K}  \\ \cline{2-7}
          & \#Itr.  & Time(Sec.) & Status &	 \#Itr.  & Time(Sec.) & Status \\ \hline 
  %		c432         &	1         &	0.105627         &	Wrong Key       &	NA        &	NA        		&	NA         			\\    \hline
  		c499         &	14        &	0.621622         &	UNSAT          &	27        &	1.12494         &	UNSAT         \\    \hline
  		c880         &	5         &	0.175454         &	Wrong Key      &	6         &	0.2497          &	UNSAT         \\    \hline
  		c1355        &	8         &	0.439975         &	UNSAT          &	14        &	1.70843         &	UNSAT         \\    \hline
  		c1908        &	3         &	0.312497         &	Wrong Key      &	2         &	0.273019        &	UNSAT          \\    \hline
  		c2670        &	9         &	0.487856         &	UNSAT          &	16        &	0.753063        &	Wrong Key       \\    \hline
  		c3540        &	8         &	0.709153         &	UNSAT          &	3         &	0.583499        &	UNSAT         \\    \hline
  		c5315        &	7         &	0.935628         &	UNSAT          &	7         &	0.975759        &	UNSAT         \\    \hline
  		c7552        &	3         &	1.20122          &	UNSAT          &	2         &	1.13674         &	UNSAT         \\    \hline
  	
       \end{tabular}}
    \end{table}

 While performing the SAT attack, we have extracted three evaluation metrics, i.e., the number of SAT iterations, SAT attack time (in seconds) and status (whether the correct key was found or not), to validate the proposed IDKLL. It can be observed from this table that SAT attack fails to determine any key ($i.e.$, ``UNSAT Model") or provides an ``wrong key" for the implementations of the proposed method. On the other hand, SAT attack identifies the correct key for all the benchmark circuits locked using exiting Anti-SAT, CAS-Lock and other SAT resistant methods in an average $2^{k}$ iterations with $2k$ keys ( here k= 8, 16). However, the SAT attack time and the number of SAT iterations are significantly low in the proposed sub-circuit replacement based IDKLL. This is because the execution of SAT attack fails or terminates unsuccessfully. Therefore, the attacker has to employ brute force only in our method. The comparison of SAT/brute-force iterations/attempts required by the proposed SubLock and existing methods are shown in Table \ref{Tab3}. Here $m$ is the number of critical inputs. The SAT attack does not fail in the existing SAT-resistant methods; it runs until the worst-case time and returns the correct key afterwards. 
  
  \begin{table*}[h!]
    \centering
     \caption{Comparison of SAT/brute-force iterations/attempts (k=2*n keys) achieved by different SAT resistant methods}
       \label{Tab3}
        {\small
       \begin{tabular}{|c|c|c|c|c|c|c|}   \hline  
         Methods & SARLock & Anti-SAT & TTLock  & CAS-Lock & Strong Anti-SAT & Proposed SubLock	 \\ \hline 
    		\#Iterations 	&	$2^k$ 	&	$2^n$ 	&	$2^k$  		&	$2^n$    & $\frac{2^n + m}{2}$ & $2^k  + (2^k)!\sum_{r=1}^{2^k-1} \frac{1}{(r)!}$ \\   \hline
  	
     \end{tabular}}
  \end{table*}

 In the proposed method, SAT attack would never identify the correct key in the proposed IDKLL method, even locking with any number of keys. Therefore, the proposed IDKLL method completely prevents the SAT attack. This is basically happening because the proposed technique uses multiple key sequences as a correct key; thus, SAT attack either eliminates all the keys or leaves with a single key sequence that can never be correct for all the input patterns. On the other hand, the existing SAT-resistant methods approach uses a single key sequence as a correct key for all input patterns. Hence, the existing SAT resistance logic locking techniques failed to prevent the SAT attack. Further, we also analyze the effectiveness of the proposed method against variants of SAT attacks such as App-SAT, removal, IFS and others. Since the proposed method is not based on a single-point function and can provide desired output corruption, applying these attacks is approximately impossible on our method. The proposed sub-circuit replacement based IDKLL method will be very effective in mitigating all the known attacks. The overhead analysis of SubLock is presented in the next subsection.
 
% Further, we also analyzed that the proposed sub-circuit replacement based IDKLL prevents the SAT-based attacks with reduced overhead compared to all the existing techniques. 
% The overhead analysis of our logic locking is presented in the next subsection.

\subsection{Overhead Results and Analysis} 
\begin{figure}[h!]    \centering
     \includegraphics[width = 0.55\columnwidth]{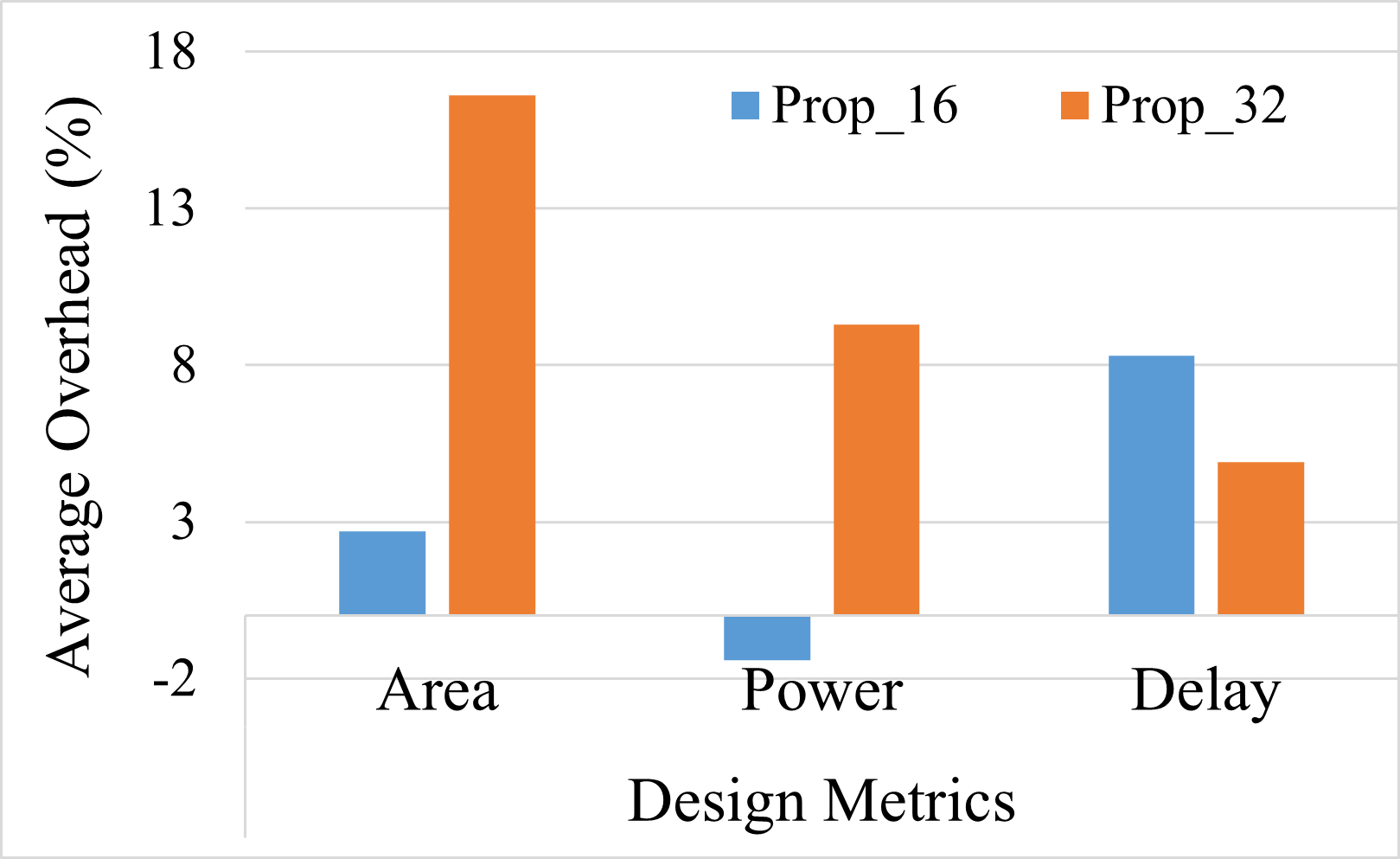}
       \caption{Average percentage overhead required by proposed method for embedding 16 and 32 keys in the small benchmarks (ISCAS-85).}
       \label{fig8_1}
  \end{figure}

  To analyze the implementation overhead of our technique, we have implemented the proposed method by embedding the varying number of keys in ISCAS and ITC benchmarks. The original, as well as locked benchmarks, are synthesized using the Cadence RTL compiler and different design metrics such as area, power and delay are extracted. It has been analyzed during synthesis that the proposed sub-circuit replacement based IDKLL (SubLock) requires a significantly large overhead for small benchmarks (ISCAS-85), even with a small key size. The average percentage overhead for the area, power and delay required by the proposed SubLock method for 16 (Prop\_32K) and 32 (Prop\_32K) keys are presented in Figure \ref{fig8_1}. Here, it can be seen that the proposed SubLock require 16\%, 9.3\% and 5\%, area, power and delay overhead for embedding 32 keys.  
    
 However, we have also evaluated the proposed SubLock method on large benchmarks with large key sizes. This evaluation proved that our SubLock method could effectively thwart SAT-based attacks with low overhead. The area, power and delay required for locking large ISCAS-89 and ITC benchmark circuits using the proposed method while embedding 32, 64, and 128 keys are shown in Table \ref{Tab1}. It can be easily observed from this table that the design metrics (area, power, delay) are not significantly increasing while changing the key size from 32 to 128. It means that the proposed SubLock method will not require high area, power, and delay while locking a design even with more large size key, i.e., K=256 or K=512. 
 
   \begin{table*}[t!]
     \centering
      \caption{Design Metrics of Proposed Sub-circuit replacement based IDKLL method for varying key-size}
       \label{Tab1}
       {\scriptsize
      \begin{tabular}{|c|c|c|c|c|c|c|c|c|c|}   \hline
       \multirow{2}{*}{Circuit} &  \multicolumn{3}{c|}{Area ($\mu m^2$)} & \multicolumn{3}{c|}{Power (nW)} &\multicolumn{3}{c|}{Delay (ps)}  \\ \cline{2-10}
       & Prop\_32K & Prop\_64K & Prop\_128K & Prop\_32K & Prop\_64K & Prop\_128K  & Prop\_32K & Prop\_64K & Prop\_128K   \\ \hline 
        	  
      	s35932 &	12371  	&	12428  	&	12673  	&	842968	&	848411 	&	884143 	&	3102 	&	3131	&	2430    \\	\hline
       	s38417 &	11826  	&	11825   &	11931   &	664726  &	695979  &	670394 	&	5855    &	5828    &	5750    \\	\hline
       	s38584 &	9936  	&	9957    &	10181   &	628548  &	633068  &	653932 	&	4152    &	4230    &	4227    \\	\hline
       	b14    &	3430    &	3574    &	3610    &	272687  &	316452  &	312861 	&	17891   &	21107   &	21448    \\	\hline
       	b15    &	6574    &	6699    &	6744    &	333576  &	345179  &	360123 	&	21749   &	21872   &	22130    \\	\hline
       	b17    &	20762   &	20826   &	20880   &	1025126 &	1036394 &	1041532	&	21228   &	20649   &	21516    \\	\hline
       	b18    &	51947   &	51994   &	52044   &	3031686 &	3060482 &	3052163 &	15119   &	14654   &	15357    \\	\hline
       	b20    &	7297    &	7318    &	7353    &	498601  &	542410  &	513114  &	23616   &	24430   &	24634    \\	\hline
       	b21    &	7547    &	7563    &	7593    &	495198  &	496952  &	531195  &	22884   &	24238   &	24256    \\	\hline
       	b22    &	10909   &	10909   &	10969   &	732907  &	748867  &	823871  &	23190   &	23444   &	23455    \\	\hline
     
     	{\bf Average} &	{\bf 14260} &	{\bf 14309} &	{\bf 14398} & {\bf 852602} & {\bf872419} & 	{\bf 884333} & 	{\bf 15879} & 	{\bf 16358 } & 	{\bf 16520}      \\       \hline
     								
    \end{tabular}}
  \end{table*}
 
  In addition, we have also implemented the existing SAT resistant methods ASB \cite{xie2016mitigating}, CAS \cite{shakya2020cas}, and SAS \cite{liu2020strong} by randomly inserting different key size blocks (K=32, K=64 and K=128) in ISACS-89 and ITC benchmarks. The same locked benchmarks are synthesized using the Cadence RTL compiler, and different design metrics, i.e., area, power and delay, are extracted. The comparison of comparative analysis of the proposed method is presented next subsection.

 \subsection{Comparative Analysis}
 To concisely compare the area, power and delay, we calculated the average of each design metric for the proposed and existing ASB \cite{xie2016mitigating}, CAS \cite{shakya2020cas}, and SAS \cite{liu2020strong} methods. The comparison of average area, power and delay required by proposed and existing methods on large ISCAS-89 and ITC benchmarks is shown in Table \ref{Tab2}. 
  
   \begin{table*}[h!]
     \centering
      \caption{Comparison of Average Design Metrics Required by Existing and Proposed Techniques}
       \label{Tab2}
       \begin{tabular}{|c|c|c|c|c|c|c|c|c|c|}   \hline
        \multirow{2}{*}{Circuit} &  \multicolumn{3}{c|}{Area ($\mu m^2$)} & \multicolumn{3}{c|}{Power (nW)} &\multicolumn{3}{c|}{Delay (ps)}  \\ \cline{2-10}
         & K=32 & K=64 & K=128 &  K=32 & K=64 & K=128  &  K=32 & K=64 & K=128   \\ \hline       	  
    	 Original    &	14254    &	14254    &	14254    &	879823    &	879823    &	879823    &	15918    &	15918    &	15918    \\	 \hline
    	 ASB    &	14339    &	14477    &	14633    &	866187    &	887057    &	906152    &	16550    &	17005    &	18098    \\	 \hline
    	 CAS    &	14350    &	14475    &	14698    &	872427    &	891093    &	916603    &	17268    &	18617    &	20898    \\	 \hline
    	 SAS    &	14396    &	14514    &	14755    &	866754    &	882127    &	912844    &	17360    &	18894    &	20379    \\	 \hline
    	 {\bf SubLock}   &	 {\bf 14260}    &	 {\bf 14309}    &	 {\bf 14398}    &	 {\bf 852602}    &	 {\bf 872419}    &	 {\bf 884333}    &	 {\bf 15879}    &	 {\bf 16358}    &	 {\bf 16520}    \\	 \hline    								
     \end{tabular}
  \end{table*}

\begin{figure}[h!]    \centering
       \includegraphics[width = 0.75\columnwidth]{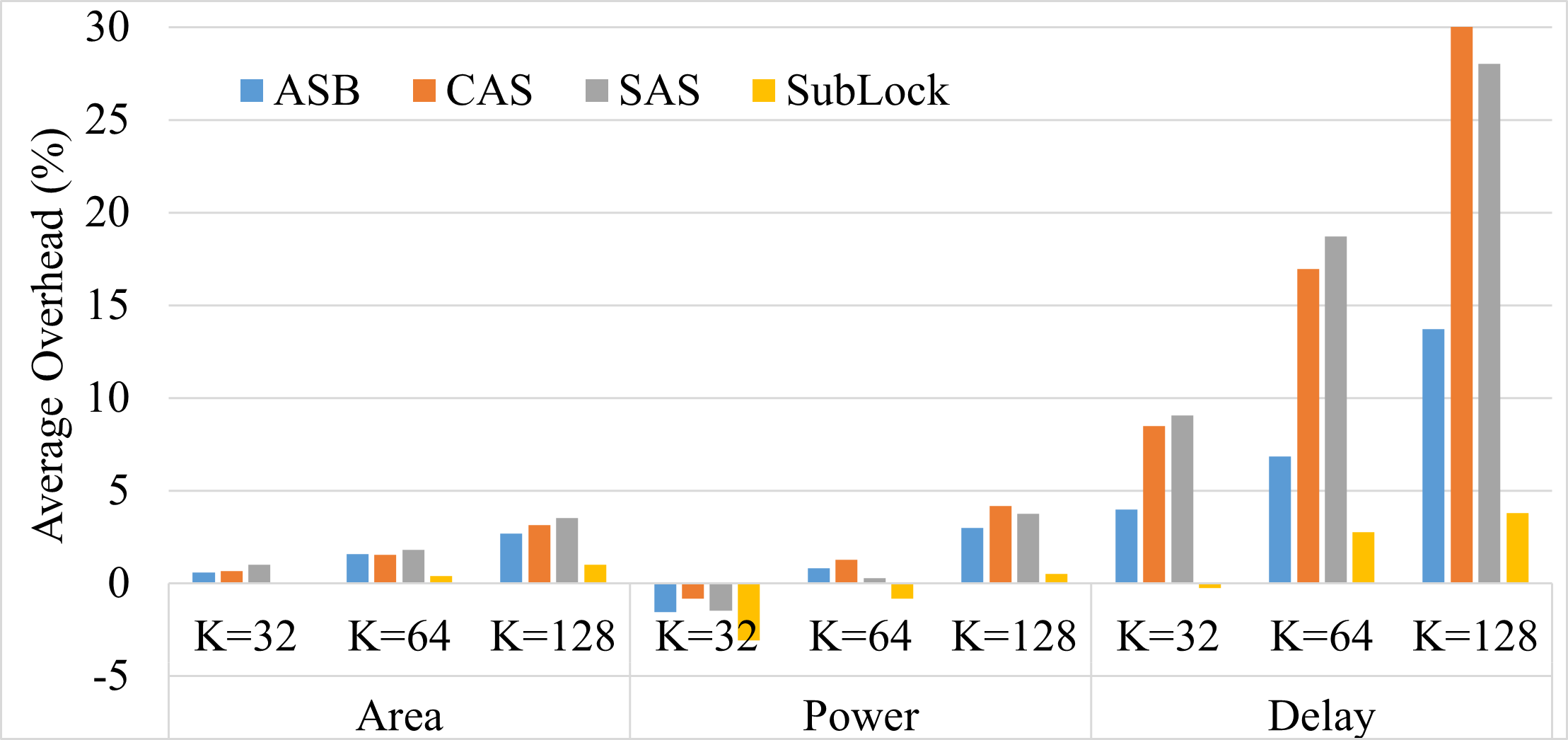}
       \caption{Comparison of average overhead (\%) required by proposed and existing methods for embedding K=16, K=32 and K=128 keys in the large ISCAS-89 and ITC benchmarks.}
        \label{fig9_1}
    \end{figure}

  It can be observed from this table that the ASB requires slightly less overhead compared to CAS and SAS blocks. However, the SAS block requires a large overhead in comparison to ASB and CAS blocks; it provides high security against SAT over ASB and CAS blocks. On the other hand, the proposed SubLock method required low area, power and delay over all these SAT-resistant methods. Furthermore, we also compute the average percentage overhead required by proposed and existing methods, as shown in Figure \ref{fig9_1}. It can be analysed from this figure that the ASB, CAS and SAS methods require low overhead for area and power, whereas they require a large overhead for the delay, specifically for K=128, i.e., more than 25\%. This is because we randomly inserted the ASB, CAS and SAS blocks in the design, causing the insertion of these blocks in critical paths. However, the delay overhead can be reduced for these methods by judicious insertion or by avoiding insertion in critical paths. On the other hand, the proposed SubLock method requires low delay overhead (i.e., 3.8\%) because we replace the small sub-circuits (3-input/4-input) with the corresponding IDKLL based locked circuit.    
 
   In addition, it can also be easily analysed that area and power overhead for all the methods are less than the 5\% even for 128-bit key size. Though the SAS block insertion-based method provides high security, it requires a large area and power overhead compared to ASB and CAS insertion-based methods. For the 128-bit key, the SAS-based method requires 3.5\% and 3.8\% area and power overhead, and the ASB method requires 2.7\% and 3\% area and power overhead receptively. All these existing methods are implemented without any obfuscation; they may be vulnerable to Removal attacks \cite{yasin2018removal}. The implementation of these methods with obfuscation will require significantly high overhead. On the other hand, the proposed SubLock method requires significantly less area and power overhead compared to ASB, CAS and SAS block insertion-based methods for all the key sizes. For the 128-bit key, the proposed method requires 1\% and 0.5\% area and power overhead, whereas, for the 64-bit key, it requires only 0.4\% and -0.8\% area and power overhead, respectively.   
   
   It can also be observed from Figure \ref{fig9_1} and Table \ref{Tab2} that the proposed method requires less power and delay (negative value in the figure for K=32 and K=64) in comparison to the original circuit. However, it is majorly happening for K=32, where both power and delay overhead are negative in the proposed method. This is happening because 1) we intentionally designed and optimized the locked sub-circuits to reduce the overhead of the proposed logic locking method, 2) the replacement of sub-circuits with corresponding IDKLL based optimized locked sub-circuits changes the original logic of the replaced sub-circuits; afterwards, the synthesizer optimizes the entire locked circuits to reduces power, area and delay. These optimizations slightly reduce the overhead of a locked circuit over the original circuit. Due to the low number of replacements, this slight reduction is clearly visualized while embedding a small number of keys (i.e., K=32 or K=64) in large benchmarks. Due to many replacements, it is not visualizing in the case of K=128. Hence, in practical key size (K=128, K=256 and K=512), the power and delay overhead in the proposed method will always be high compared to the original circuit.

\begin{figure}[h!]    \centering
      \includegraphics[width = 0.7\columnwidth]{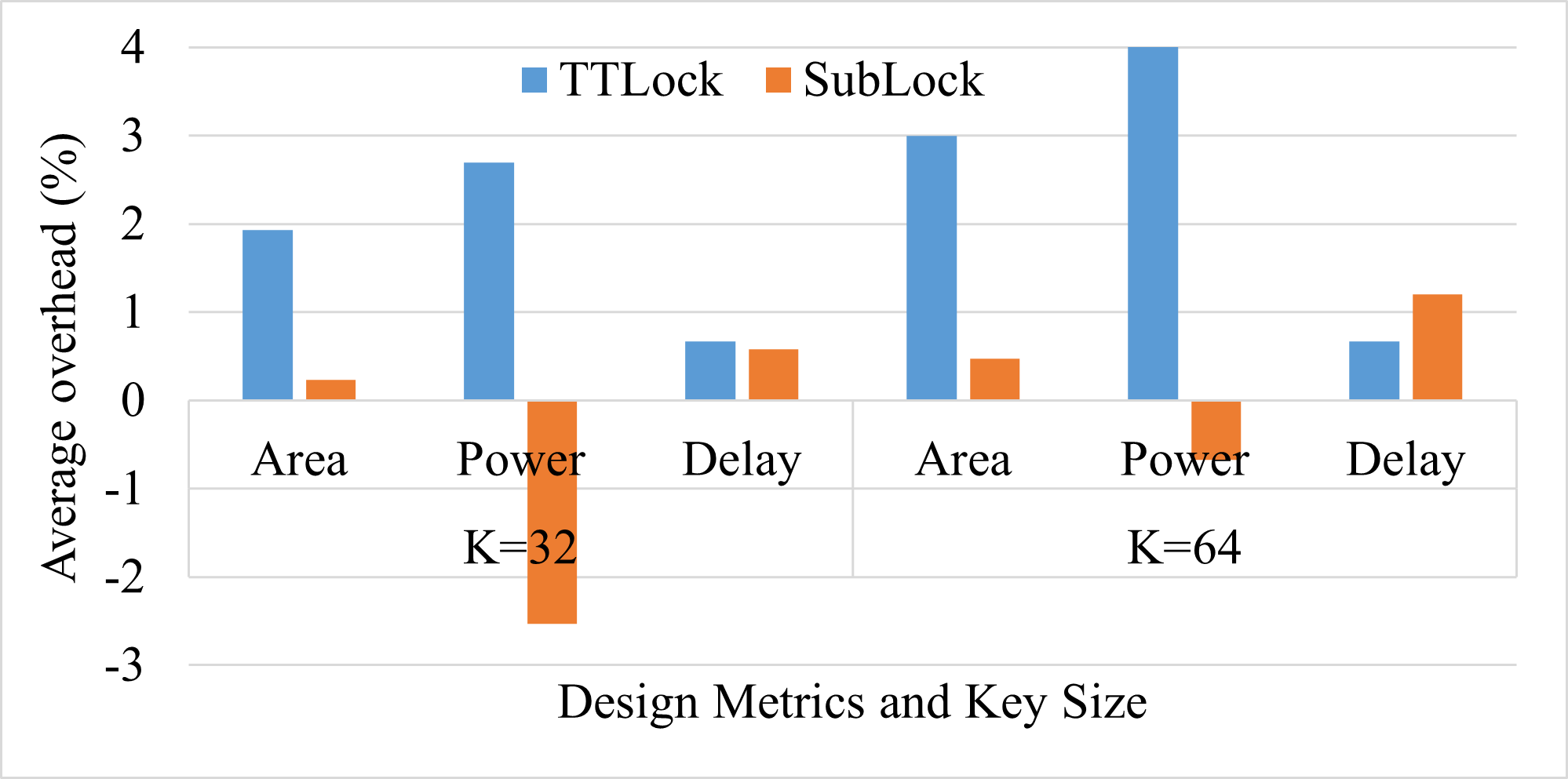}
       \caption{Comparison of average percentage overhead of proposed SubLock and TTLock methods on ISCAS-89 benchmarks for 32 and 64 keys.}
       \label{fig8}
    \end{figure}

  Besides the above, we have also compared our approach with the TTLock \cite{yasin2017lock}. The comparison of the average percentage overhead required by proposed and TTLock methods on large ISACS-89 benchmarks for the key size of 32 and 64 is shown in Figure \ref{fig8}. It can be observed here that the delay overhead in TLock is approximately near to our approach. But TTLock requires significantly high area and power overhead compared to the proposed approach. We have also analyzed the implementation overhead of the SFLL \cite{yasin2017provably} method. It is found that the proposed method also requires significantly less overhead compared to SFLL. The SFLL is the extension of TTLock, which provides high security over TTLock but requires more overhead over the TTLock. Since the proposed SubLock method requires less overhead over TTLock even for the same key size, thus the proposed method will also require low overhead over the SFLL method for the same key size. In addition to these methods, other recently reported methods such as CORALL \cite{9070188}, DisORC \cite{9214869}, G-Anti-SAT \cite{zhou2021generalized} and SKG-Lock+  \cite{nguyen2022skg} are also required around 10\% overhead while only increasing the security against SAT attack not preventing it.

   In the above analysis, we compared and analyzed that the implementation of the proposed SubLock method for the same key size requires low implementation overhead compared to existing SAT-resistant methods. On the other hand, it can be observed from Eq. (\ref{eq9}) that the proposed method achieves minimum $2^k$ more attack iterations for the same key size over the existing ASB, CAS, and SAS methods. It means the proposed method can achieve the same security with the 64-bit key, which is achieved by existing methods with a 128-bit key. In this case, the required overhead of the proposed method will be further reduced while comparing the proposed SubLock and existing methods for the same security. For the 64-bit key, the proposed SubLock method can provide the same or higher security with 0.4\%, -0.8\% and 2.76\% area, power, and delay overhead. In contrast, the existing SAS/CAS method requires around 3.5\%/3\%, 3.8\%/4.2\% and 28\%/31\% area, power and delay overhead, respectively for achieving the same security. Here, it is clear that the proposed method requires significantly less overhead than all the existing methods for the same security. Overall, it can be easily said that the proposed SubLock method outperforms all the well-known SAT-resistant methods in terms of security and design overhead.

\section{Conclusion} \label{con}
  This paper proposes a new SAT-resistant logic locking method based on the idea of input dependent key-based logic locking called IDKLL. The existing SAT-resistant logic locking techniques utilize a single key sequence as a correct key to unlock the design, whereas the proposed IDKLL approach uses multiple key sequences as correct to unlock the design functionality for all inputs. The use of multiple key sequences as the correct key neutralizes the SAT attack completely. In order to lock the whole design functionality using IDKLL, we propose a lightweight sub-circuit replacement based IDKLL method called SubLock that provides effective security against SAT based attacks with low overhead. The proposed SubLock approach replaces the original small sub-circuitries with their corresponding IDKLL based locked circuits. Further, we present the security analysis of the proposed method, which shows that our method tremendously increases the security or brute force attempts over all the existing SAT-resistant methods and effectively prevents SAT based attacks. The experimental evaluation of the proposed SubLock on ISCAS and ITC benchmarks shows that the proposed method effectively prevents the SAT attack while requiring very low overhead compared to the well-known method such as Anti-SAT, CAS-Lock and Strong Anti-SAT.  

  The concept of IDKLL is based on using multiple correct key sequences to unlock the design for all inputs. This paper uses LUT based tamper-proof memory to store multiple valid key sequences and retrieve the correct key sequence corresponding to a given input set. Implementing LUT based tamper-proof memory for storing multiple key sequences may require high implementation costs over implementing normal taper-proof memory for storing a single key sequence. Therefore, our future work will focus on identifying the cost-effective implementation/way to store multiple correct key sequences.

\bibliographystyle{unsrtnat}
\bibliography{template} 

\vspace{15pt}
%\begin{IEEEbiography}
%{\includegraphics[width=1in,height=1.25in,clip,keepaspectratio]{figures/Vijay_Pic}}
{\includegraphics[width=1in,height=1.25in,clip,keepaspectratio]{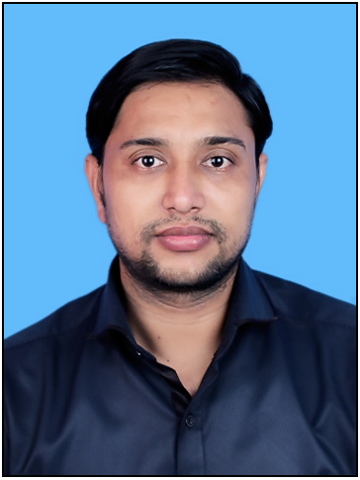}}
{Vijaypal Singh Rathor}
%received his M.Tech. in Information Security from Maulana Azad National Institute of Technology, Bhopal, and Ph.D. from ABV-Indian Institute of Information Technology and Management Gwalior, India in 2014 and 2020 respectively. 
 is currently working as an Assistant Professor in the Department of CSE at PDPM Indian Institute of Information Technology, Design and Manufacturing (IIITDM), Jabalpur, India, since August 2021. %Before joining IIITDM, he worked as an Assistant Professor at Thapar Institute of Engineering and Technology, Patiala, and Bennett University, Greater Noida, India. 
%He has over three years of teaching experience to his credit. He has published over 10 research articles in International Journals/Conferences of repute. 
His research interest includes Hardware Security, Machine Learning and IoT, and Cloud Computing. He is also a member of IEEE since 2017. 

%\end{IEEEbiography}

\vspace{20pt}
%\begin{IEEEbiography}
%{\includegraphics[width=1.1in,height=1.25in,clip,keepaspectratio]{figures/Munesh_Pic}}
{\includegraphics[width=1.1in,height=1.25in,clip,keepaspectratio]{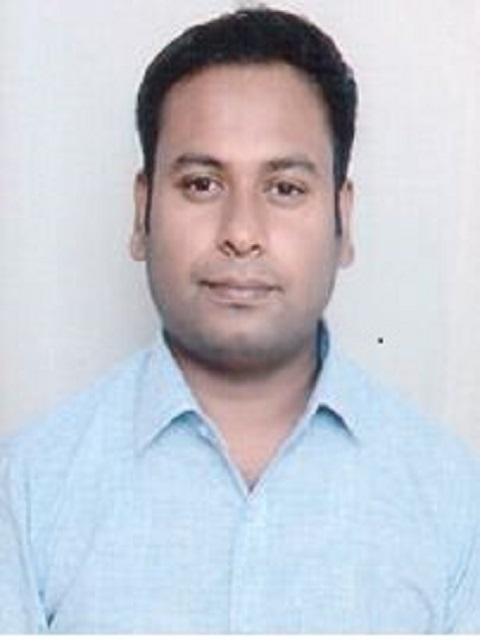}}
{Munesh Singh}
%received the Ph.D. degree in Computer Science and Engineering from National Institute of Technology (NIT), Rourkela, India, in 2017. He has industrial experience of one year with the Honeywell Technology Solution Lab, Bangalore, India. He 
is currently working as an Assistant Professor in PDPM Indian Institute of Information Technology, Design and Manufacturing (IIITDM), Jabalpur, India. %He was previously working as an Assistant Professor in IIITDM, Kancheepuram, India. 
He has published many research articles in IEEE, Springer, and Elsevier journals.
%He works as a reviewer for Wiley, IEEE, and Springer.
 His research interests include Sensor  Networks, Cooperative Computing, Radar Surveillance, Cyber Security, Machine Learning, and Intelligent Robotics. %He is a member of IEEE and IAENG.
 
%\end{IEEEbiography}

%\vspace{20pt}
%\begin{IEEEbiography}
%{\includegraphics[width=1in,height=1.25in,clip,keepaspectratio]{figures/kshira.png}}
{\includegraphics[width=1in,height=1.25in,clip,keepaspectratio]{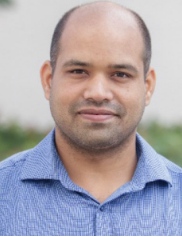}}
{Kshira Sagar Sahoo} (Senior Member, IEEE)
completed his Ph.D. (2019) and M.Tech (2014) from Dept. of CSE, National Institute of Technology, Rourkela, India and Indian Institute of Technology, Kharagpur, India respectively. He is currently acting as a postdoctoral \textit{Kempe} fellow at the dept of computing science, Ume\r{a} University, Sweden. 
He has authored more than 90 research articles in top tier journals and conferences, 2 edited books, and 8 granted and pending patents. 
He is recognized
as top 2\% scientists in the world 2023 list by Stanford University. 
His research interests are SDN, SDIoT, Edge Computing, IoT, Machine Learning and Distibuted Computing. He is a senior member of IEEE and IETE.

%\end{IEEEbiography}

\vspace{20pt}
% \begin{IEEEbiography}
% [{\includegraphics[height=1.25in, keepaspectratio]{Saraju_Big}}]
% {Saraju P. Mohanty} (Senior Member, IEEE) received the bachelor's degree (Honors) in electrical engineering from the Orissa University of Agriculture and Technology, Bhubaneswar, in 1995, the master’s degree in Systems Science and Automation from the Indian Institute of Science, Bengaluru, in 1999, and the Ph.D. degree in Computer Science and Engineering from the University of South Florida, Tampa, in 2003. He is a Professor with the University of North Texas. His research is in ``Smart Electronic Systems'' which has been funded by National Science Foundations (NSF), Semiconductor Research Corporation (SRC), U.S. Air Force, IUSSTF, and Mission Innovation. He has authored 450 research articles, 5 books, and invented 9 granted/pending patents. His Google Scholar h-index is 51 and i10-index is 219 with 11,000 citations. He is a recipient of 16 best paper awards, Fulbright Specialist Award in 2020, IEEE Consumer Electronics Society Outstanding Service Award in 2020, the IEEE-CS-TCVLSI Distinguished Leadership Award in 2018, and the PROSE Award for Best Textbook in Physical Sciences and Mathematics category in 2016. He has delivered 18 keynotes and served on 14 panels at various International Conferences. He has been the Editor-in-Chief of the IEEE Consumer Electronics Magazine during 2016-2021 and currently serves on the editorial board of 8 journals/transactions.
% \end{IEEEbiography}

%\begin{IEEEbiography}
%{\includegraphics[height=1.25in, keepaspectratio]{figures/Saraju_Big.PNG}}
{\includegraphics[height=1.25in, keepaspectratio]{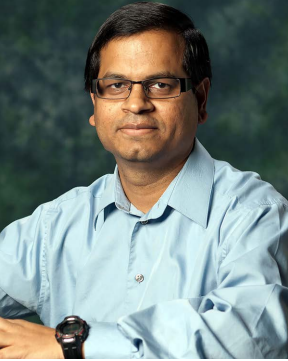}}
{Saraju P. Mohanty} (Senior Member, IEEE) received the bachelor’s degree (Honors) in electrical engineering from the Orissa University of Agriculture and Technology, Bhubaneswar, in 1995, the master’s degree in Systems Science and Automation from the Indian Institute of Science, Bengaluru, in 1999, and the Ph.D. degree in Computer Science and Engineering from the University of South Florida, Tampa, in 2003. He is a Professor with the University of North Texas. His research is in ``Smart Electronic Systems’’ which has been funded by National Science Foundations (NSF), Semiconductor Research Corporation (SRC), U.S. Air Force, IUSSTF, and Mission Innovation. He has authored 500 research articles, 5 books, and 10 granted and pending patents. His Google Scholar h-index is 56 and i10-index is 258 with 14,000 citations. He is regarded as a visionary researcher on Smart Cities technology in which his research deals with security and energy aware, and AI/ML-integrated smart components. He introduced the Secure Digital Camera (SDC) in 2004 with built-in security features designed using Hardware Assisted Security (HAS) or Security by Design (SbD) principle. He is widely credited as the designer for the first digital watermarking chip in 2004 and first the low-power digital watermarking chip in 2006. He is a recipient of 19 best paper awards, Fulbright Specialist Award in 2021, IEEE Consumer Electronics Society Outstanding Service Award in 2020, the IEEE-CS-TCVLSI Distinguished Leadership Award in 2018, and the PROSE Award for Best Textbook in Physical Sciences and Mathematics category in 2016. He has delivered 29 keynotes and served on 15 panels at various International Conferences. He has been serving on the editorial board of several peer-reviewed international transactions/journals, including IEEE Transactions on Big Data (TBD), IEEE Transactions on Computer-Aided Design of Integrated Circuits and Systems (TCAD), IEEE Transactions on Consumer Electronics (TCE), and ACM Journal on Emerging Technologies in Computing Systems (JETC). He has been the Editor-in-Chief (EiC) of the IEEE Consumer Electronics Magazine (MCE) during 2016-2021. He served as the Chair of Technical Committee on Very Large Scale Integration (TCVLSI), IEEE Computer Society (IEEE-CS) during 2014-2018 and on the Board of Governors of the IEEE Consumer Electronics Society during 2019-2021. He serves on the steering, organizing, and program committees of several international conferences. He is the steering committee chair/vice-chair for the IEEE International Symposium on Smart Electronic Systems (IEEE-iSES), the IEEE-CS Symposium on VLSI (ISVLSI), and the OITS International Conference on Information Technology (OCIT). He has mentored 3 post-doctoral researchers, and supervised 17 Ph.D. dissertations, 27 M.S. theses, and 27 undergraduate projects.
%\end{IEEEbiography}

%%% Uncomment this line and comment out the ``thebibliography'' section below to use the external .bib file (using bibtex) .





% \end{thebibliography}

\end{document}